\documentclass[12pt]{article}
\usepackage{amssymb,amssymb, latexsym}
\usepackage{amsfonts,psfrag,amsmath,bbm,color,multirow}
\usepackage{placeins} 
\usepackage{graphicx}
\usepackage{epstopdf}
\graphicspath{{./Source_Sec14/}}

\newtheorem{remark}{Remark}[section]

\def\PP{{{\rm l}\kern - .15em {\rm P} }}
\def\PN2{{\PP_{N}-\PP_{N-2}}}

\newcommand{\bphi}{\boldsymbol{\varphi}}

\newcommand{\bu}{\boldsymbol{u}}
\newcommand{\bU}{\boldsymbol{U}}

\newcommand{\bw}{\boldsymbol{w}}

\newcommand{\bx}{\boldsymbol{x}}
\newcommand{\bX}{\boldsymbol{X}}

\newcommand{\tA}{\tilde{A}}

\newcommand{\oa}{\overline{a}}

\newcommand{\obu}{\overline{\boldsymbol u}}

\newcommand{\deleted}[1]{{}}

\begin{document}

\title{An Evolve-Then-Filter Regularized Reduced Order Model \\ For Convection-Dominated Flows}

%
\author{David Wells, Zhu Wang, Xuping Xie, Traian Iliescu}

\maketitle

\begin{abstract}
In this paper, we propose a new evolve-then-filter reduced order model (EF-ROM).
This is a regularized ROM (Reg-ROM), which aims at the numerical stabilization of proper orthogonal decomposition (POD) ROMs for convection-dominated flows.
We also consider the Leray ROM (L-ROM).
These two Reg-ROMs use explicit ROM spatial filtering to smooth (regularize) various terms in the ROMs. 
Two spatial filters are used: a POD projection onto a POD subspace (Proj) and a new POD differential filter (DF). 
The four Reg-ROM/filter combinations are tested in the numerical simulation of the three-dimensional flow past a circular cylinder at a Reynolds number $Re=1000$. 
Overall, the most accurate Reg-ROM/filter combination is EF-ROM-DF.  
Furthermore, the spatial filter has a higher impact on the Reg-ROM than the regularization used.
Indeed, the DF generally yields better results than Proj for both the EF-ROM and L-ROM.
Finally, the CPU times of the four Reg-ROM/filter combinations are orders of magnitude lower than the CPU time of the DNS.
\end{abstract}



\medskip
{\bf Key Words:} Reduced order model, proper orthogonal decomposition, regularized model, stabilization method, spatial filter, differential filter.

\medskip
{\bf Mathematics Subject Classifications (2000)}: 65M15, 65M60

\vspace{-6pt}

\section{Introduction}
\vspace{-2pt}

Important applications require repeated numerical simulations with a high computational cost~\cite{hesthaven2015certified,HLB96,noack2011reduced,quarteroni2015reduced}. 
To enforce a balance between accuracy and computational cost, reduced order models (ROMs) emerge as natural choices. 
The proper orthogonal decomposition (POD) is one of the most successful approaches for ROM development.  
An accurate numerical simulation is used in POD to extract the dominant structures, which are then used in a Galerkin approximation of the underlying equations~\cite{HLB96,Sir87abc}. 
In this paper, POD will be exclusively used to construct ROMs.

Although ROMs have been successful across a range of disciplines, their use in {\it convection-dominated flows} has been hampered by their notorious {\it numerical instability}.
Indeed, since the number of POD basis functions is usually small (e.g., $ \mathcal{O}(10)$), the standard ROM provides an efficient surrogate model for the underlying flows. 
However, for convection-dominated flows, the standard ROM does not represent a viable tool~\cite{AHLS88,balajewicz2013low,ballarin2015supremizer,giere2015supg,pacciarini2014stabilized,quarteroni2011certified,wang2011two,wang2012proper}, since it generally yields spurious numerical oscillations.
These results clearly indicate the need for {\it ROM stabilization}. 
Over the years, numerous strategies have been devised to alleviate this numerical instability of the standard ROM (see, e.g.,~\cite{amsallem2012stabilization,AHLS88,balajewicz2013low,ballarin2015supremizer,barone2009stable,bergmann2009enablers,carlberg2013gnat,kalashnikova2010stability,quarteroni2011certified,wang2011two,wang2012proper,xiao2014non} and references therein).

In this paper, we investigate {\it regularized ROMs (Reg-ROMs)}, which use {\it explicit spatial filtering} to stabilize the ROMs and enable their use in convection-dominated flows.
The idea of using a spatial filter to stabilize numerical simulations with spectral methods has a long history~\cite{boyd2001chebyshev, CHQZ88, FM01, MF99, pasquetti2002comments}.
In ROMs, spatial filtering has been used as a preprocessing step, to filter out the noise in the snapshot data, i.e., in the generation of the POD modes (see, e.g., Section 5 in~\cite{aradag2011filtered} for a survey of relevant work).
We emphasize, however, that our approach is {\it fundamentally different}.
Indeed, we {\it explicitly} use spatial filters to smooth (regularize) different terms in the {\it actual ROMs}, i.e., we modify the mathematical model, not the input data.
The first explicit ROM spatial filter that we use is the POD projection (Proj), which is the projection of the ROM variables on a subspace of the POD space~\cite{KV01,wang2011two,wang2012proper}.
The other spatial filter that we employ is a POD differential filter (DF)~\cite{germano1986differential-b,germano1986differential,sabetghadam2012alpha}.

In this paper, we put forth a new evolve-then-filter Reg-ROM.
We also extend and investigate the Leray ROM proposed in~\cite{sabetghadam2012alpha}.
To our knowledge, the new EF-ROM and the L-ROM are the only Reg-ROMs in current use.
Note that these Reg-ROMs are fundamentally different from the calibration approaches used in~\cite{abou2016new,alekseev2001analysis,cordier2010calibration,wang20152d,weller2009robust}.
Indeed, the Reg-ROMs investigated in this paper use equations that are different from those employed in the standard ROM. 
The calibration models in~\cite{abou2016new,alekseev2001analysis,cordier2010calibration,wang20152d,weller2009robust}, on the other hand, utilize the standard ROM equations and only calibrate their coefficients by utilizing a Tikhonov type regularization. 

The rest of the paper is organized as follows: 
In Section~\ref{sec:rom}, we briefly present the POD and standard ROM. 
In Section~\ref{sec:rom-spatial-filtering}, we discuss the ROM spatial filters (i.e., the Proj and DF). 
In Section~\ref{sec:reg-roms}, we employ the Proj and DF to lay foundations for the new evolve-then-filter Reg-ROM and the Leray Reg-ROM.
In Section~\ref{sec:numerical-tests}, we present numerical results for the two Reg-ROMs and two ROM spatial filters for the three-dimensional (3D) flow past a circular cylinder at a Reynolds number $Re=1000$. 
Finally, we draw conclusions in Section~\ref{sec:conclusions}.

\vspace{-6pt}

\section{Reduced Order Modeling}
    \label{sec:rom}

    In this section, the POD, the standard Galerkin ROM and
    the centering trajectory are briefly presented. The {\it Navier-Stokes
    equations (NSE)} are used as mathematical model:
    \begin{eqnarray}
    && \bu_t
    - Re^{-1} \Delta \bu
    + \bu \cdot \nabla \bu
    + \nabla p
    = {\bf 0} \, ,
    \label{eqn:nse-1}                                                         \\
    && \nabla \cdot \bu
    = 0 \, ,
    \label{eqn:nse-2}
    \end{eqnarray}
    where $\bu$ is the velocity, $p$ the pressure and $Re$
    the Reynolds number. 
In this paper, the NSE~\eqref{eqn:nse-1}--\eqref{eqn:nse-2} are supplemented with the initial condition $\bu(\bx,0) = \bu_0(\bx)$ and steady Dirichlet boundary conditions.

    \subsection{Proper Orthogonal Decomposition (POD)}
        \label{sec:pod}
        One of the most popular reduced order modeling techniques is the
        POD, which we briefly describe next.  For more details, the reader is referred to, e.g.,~\cite{HLB96,noack2011reduced,Sir87abc}. The POD starts with the snapshots
        $\{\bu^1_h,\ldots, \bu^{N_s}_h\}$, which are numerical approximations of
        \eqref{eqn:nse-1}--\eqref{eqn:nse-2} at $N_s$ different time instances.  
The finite element (FE) solutions of~\eqref{eqn:nse-1}--\eqref{eqn:nse-2} are considered as snapshots in this section. 
We emphasize, however, that other numerical methods can be used instead.
The POD seeks a low-dimensional basis that approximates the snapshots
        optimally with respect to a certain norm. In this paper, the commonly
        used $L^2$-norm will be chosen. The solution of the minimization
        problem is equivalent to the solution of the eigenvalue problem
        \begin{equation}
            \label{eq:pod_ev}
            YY^\top M \bphi_j = \lambda_j \bphi_j,
            \quad j=1,\ldots,N,
        \end{equation}
        where $\bphi_j$ and $\lambda_j$ denote the vector of the FE coefficients
        of the POD basis functions and the POD eigenvalues, respectively, $Y$
        denotes the snapshot matrix, whose columns correspond to the FE
        coefficients of the snapshots, $M$ denotes the FE mass matrix, and $N$
        is the dimension of the FE space $\bX^h$~\cite{caiazzo2013numerical}.
        The eigenvalues are real and non-negative, so they can be ordered as
        follows:
        \begin{equation}
            \lambda_1 \ge \lambda_2 \ge \ldots \ge \lambda_R \ge \lambda_{R + 1}
            = \ldots = \lambda_N = 0.
        \end{equation}
        The POD basis consists of the normalized functions $\{
        \bphi_{j}\}_{j=1}^{r}$, which correspond to the first $r\le N$ largest
        eigenvalues. Thus, the POD space is defined as $\bX^r := \text{span} \{
        \bphi_1, \ldots, \bphi_r \}$.

The centering trajectory method is popular in ROM development \cite{HLB96}.
In this approach, the snapshots $\{ \bu_h^1, \ldots, \bu_h^{N_s} \}$ are replaced by $\{ \bu_h^1 - \bU , \ldots, \bu_h^{N_s} - \bU \}$, where $\bU = \frac{1}{N_s} \sum_{i=1}^{N_s} \bu_h^i$ is the centering trajectory and $\bu_h^i - \bU$ are the snapshot fluctuations. 
Thus, the POD basis functions are computed from the snapshot fluctuations $\bu_h^i - \bU, \ i = 1, \ldots, N_s$. 
The centering trajectory method is especially useful for problems that have steady-state nonhomogeneous Dirichlet boundary conditions, such as the boundary conditions in the 3D flow past a cylinder investigated in Section~\ref{sec:numerical-tests}. 
Since in this case the snapshot fluctuations $\bu_h^i - \bU, \ i = 1, \ldots, N_s$ satisfy homogeneous Dirichlet boundary conditions, the centering trajectory method avoids the challenges posed by the nonhomogeneous boundary conditions in ROMs (see, however, \cite{graham1999optimal} for alternative approaches).
Thus, in what follows, we will use the centering trajectory approach.

\subsection{The Galerkin ROM (G-ROM)}
    \label{sec:g-rom}
    
To develop the standard Galerkin ROM, we start by considering the POD approximation of the velocity 
\begin{equation}    
	{\bu}_r(\bx,t) 
	\equiv \bU(\bx) + \sum_{j=1}^r a_j(t) \bphi_j(\bx),
	\label{eqn:g-rom-1}
\end{equation} 
where $\{a_{j}(t)\}_{j=1}^{r}$ are the sought time-varying coefficients that represent the POD-Galerkin trajectories. 
By using the POD basis in a Galerkin approximation of the NSE, the standard {\it Galerkin ROM (G-ROM)} is obtained: $\forall \, i = 1, \ldots, r,$
\begin{eqnarray}
	\left( \frac{\partial \bu_{r}}{\partial t} , \bphi_i \right)
	+ \frac{2}{Re} \, \biggl( \mathbbm{D}(\bu_{r}) , \nabla \bphi_i \biggr)
	+ \biggl( (\bu_{r} \cdot \nabla) \bu_{r} , \bphi_i \biggr)
	= 0 \, ,
\label{eqn:g-rom-2}
\end{eqnarray}
where
$\mathbbm{D}(\bu_{r}) := (\nabla \bu^r + (\nabla \bu^r)^\top) / 2$ is the deformation tensor  of $\bu_{r}$.
The G-ROM~\eqref{eqn:g-rom-2} yields the following autonomous dynamical system for the vector of time coefficients, ${\bf a}(t)$:
\begin{equation}
  	\dot{\bf a} = {\bf b} + {\bf A} {\bf a}   + {\bf a}^\top {\bf B} {\bf a} ,
	\label{eqn:g-rom-3}
\end{equation}
\noindent 
where $\bf{b}$, $\bf{A}$, and $\bf{B}$ correspond to the constant, linear, and quadratic terms in the numerical discretization of the NSE~\eqref{eqn:nse-1}--\eqref{eqn:nse-2}, respectively. 
The initial conditions are obtained by projection:
\begin{equation}
	a_j(0)
	= \left(
		{\bu}_0 - {\bf U} , \bphi_j
		\right), \quad j=1, \ldots, r.
	\label{eqn:g-rom-4}
\end{equation}
The finite dimensional system \eqref{eqn:g-rom-3} can be written componentwise  as follows:
For all $i = 1, \ldots, r$,
\begin{eqnarray}
	\dot{a}_i(t) 
	=  b_i
	+ \sum_{m=1}^{r} A_{im}a_m(t) + \sum_{m=1}^r \sum_{n=1}^r B_{imn}a_n(t)a_m(t),
	\label{eqn:g-rom-5}
\end{eqnarray}
where
\begin{eqnarray}
	&& \hspace*{-1.3cm} 
	b_i 
	= -\bigl( {\bf U} \cdot \nabla {\bf U} , \bphi_i \bigr) 
	- \frac{2}{Re} \left( \frac{\nabla {\bf U} +\nabla {\bf U}^{\top}}{2} , \nabla \bphi_i \right), 
	\label{eqn:g-rom-6}\\
	&& \hspace*{-1.3cm} 
	A_{im} 
	= - \bigl( {\bf U} \cdot \nabla \bphi_m , \bphi_i \bigr) 
	- \bigl(\bphi_m \cdot \nabla {{\bf U}} , \bphi_i \bigr) 
	- \frac{2}{Re} \left( \frac{\nabla \bphi_m 
	+ \nabla {\bphi_m}^{\top}}{2} , \nabla \bphi_i \right),
	\label{eqn:g-rom-7} \\
	&& \hspace*{-1.3cm} 
	B_{imn} 
	= - \bigl( \bphi_m \cdot \nabla \bphi_n , \bphi_i \bigr) \, . 	
	\label{eqn:g-rom-8}
\end{eqnarray}

\section{Explicit ROM Spatial Filtering}
    \label{sec:rom-spatial-filtering}

The Reg-ROMs discussed in Section~\ref{sec:reg-roms} use explicit ROM spatial filtering to smooth the flow variables and increase the numerical stability of the models. 
In spectral methods, spatial filtering has been used for a long time to stabilize numerical simulations of convection-dominated flows~\cite{boyd2001chebyshev, CHQZ88, FM01, MF99, pasquetti2002comments}.
In ROMs, spatial filtering has been used as a {\it preprocessing} step, to filter out the noise in the snapshot data and, thus, in the generation of the POD modes (see, e.g., Section 5 in~\cite{aradag2011filtered} for a survey of relevant work).
We emphasize, however, that the spatial filtering used in Reg-ROMs is fundamentally different. 
Indeed, in the preprocessing spatial filtering used in~\cite{aradag2011filtered}, the snapshots (i.e., the input data) are spatially filtered, but the ROMs (i.e., the mathematical models) are not.
In Reg-ROMs, on the other hand, some of the ROM terms are explicitly filtered and, thus, the mathematical model is modified.
Two types of explicit ROM spatial filters are considered in this section: a POD projection (Section~\ref{sec:pod-projection}) and a POD differential filter (Section~\ref{sec:differential-filter}). 
The properties of these explicit ROM spatial filters are discussed in Section~\ref{sec:pod-spatial-filter-properties}.

    \subsection{The POD Projection (Proj)}
        \label{sec:pod-projection}
        For a fixed $r_1 < r$ and a given $\bu^r \in \bX^r$, the {\it POD projection (Proj)}
        seeks $\obu^r \in \bX^{r_1}$ such that
        \begin{eqnarray}
            \label{eqn:pod-projection}
            \left( \obu^r, \bphi_j \right)
            = ( \bu^r , \bphi_j ),
            \quad \forall \, j=1, \ldots r_1 .
        \end{eqnarray}
        The Proj~\eqref{eqn:pod-projection} has been used for
        theoretical purposes, e.g., in the error analysis of the
        G-ROM~\cite{KV01}. 
To our knowledge, the Proj has been used as an explicit ROM spatial filter only in~\cite{wang2012proper}.

        \subsection{The POD Differential Filter (DF)}
            \label{sec:differential-filter}

            The {\it POD differential filter (DF)} is defined as follows: Let $\delta$ be
            the radius of the DF. For a given $\bu^r \in
            \bX^r$, find $\obu^r \in \bX^r$ such that
            \begin{eqnarray}
                \label{eqn:pod-differential-filter}
                \biggl(
                    \left(
                        I - \delta^2 \Delta
                    \right) \obu^r , \bphi_j
                \biggr)
                = (\bu^r, \bphi_j),
                \quad \forall \, j=1, \ldots r \, .
            \end{eqnarray}
Differential filters have been used in the simulation of convection-dominated flows with standard numerical methods~\cite{germano1986differential-b,germano1986differential}. 
In reduced order modeling, the DF was first used in~\cite{sabetghadam2012alpha} in a periodic, one-dimensional (1D) setting.
In this paper, we extend the DF used in~\cite{sabetghadam2012alpha} to a non-periodic, 3D setting.

\subsection{ROM Spatial Filter Properties}
    \label{sec:pod-spatial-filter-properties}

The Proj~\eqref{eqn:pod-projection} and DF~\eqref{eqn:pod-differential-filter} share several appealing properties~\cite{BIL05}:
(i) They act as spatial filters, since they eliminate the small scales (i.e.,
    high frequencies) from the input. Indeed, the DF~\eqref{eqn:pod-differential-filter} uses an elliptic operator to
    smooth the input variable. The effect of the Proj on an input from
    the POD space $\bX^r$ is the elimination of the components corresponding to
    the POD basis functions $\{ \bphi_{j} \}_{j=r_1+1}^{r}$, which generally
    correspond to small spatial scales (high frequencies).
(ii) Both ROM spatial filters have a low computational overhead.
    Indeed, the Proj amounts to simply setting $a_j=0, \, j = r_1+1,
    \ldots, r$ in the expansion ${\bu}_r({\bx},t) = \bU(\bx) + \sum_{j=1}^r a_j(t)
    \bphi_j({\bx})$. The DF~\eqref{eqn:pod-differential-filter}, on the other hand, amounts to
    solving a linear system with a very small $r \times r$ matrix that is
    precomputed.
(iii) Both ROM spatial filters preserve incompressibility in the NSE, since they
    are linear operators.

    Although the Proj~\eqref{eqn:pod-projection} and the DF~\eqref{eqn:pod-differential-filter} share the above
    properties, they are different in one significant aspect. 
Indeed, both the DF and Proj are {\it explicit ROM spatial filters} that eliminate the small spatial scales from the input data.
We emphasize, however, that the DF uses an {\it explicit length scale} $\delta$ (i.e., the radius of the filter), whereas Proj employs an {\it implicit length scale}, which is determined by the collection of POD modes $\{ \bphi_j \}_{j = r_1+1}^{r}$. 

\section{Regularized ROMs (Reg-ROMs)}
    \label{sec:reg-roms}

	In the ROM setting, to ensure a low
    computational cost, only relatively few POD modes are generally
    included. Using the standard G-ROM~\eqref{eqn:g-rom-2} with a few POD
    modes in realistic flows yields numerical
    oscillations~\cite{giere2015supg,wang2011two,wang2012proper}.
To alleviate this unstable behavior, we propose a new Reg-ROM, the evolve-then-filter Reg-ROM. 
Furthermore, we extend to the 3D, non-periodic NSE the Reg-ROM proposed in~\cite{sabetghadam2012alpha}.
These two Reg-ROMs use the spatial filters defined in Section~\ref{sec:rom-spatial-filtering} to increase their numerical stability. 
Since the computational cost of the spatial filters is negligible, the computational cost of these Reg-ROMs will be very low, similar to the computational cost of the G-ROM.

\subsection{The Leray ROM (L-ROM)}
	\label{sec:l-rom}

We extend to the 3D, non-periodic NSE the Leray regularized ROM (L-ROM) proposed in~\cite{sabetghadam2012alpha} for the 1D, periodic Kuramoto-Sivashinsky equations: $\forall \, n = 0, \ldots, M$ and $\forall \, i = 1, \ldots, r,$
    \begin{eqnarray}
        \left(
            \frac{\bu_{r}^{n+1} - \bu_{r}^{n}}{\Delta t} , \bphi_{i}
        \right)
        + \frac{2}{Re} \, \biggl(
            \mathbbm{D}(\bu_r^n) ,
            \nabla \bphi_{i}
        \biggr)
        + \biggl(
            (\obu_r^n \cdot \nabla) \bu_r^n ,
            \bphi_{i}
        \biggr)
        = 0 \, ,
    \label{eqn:l-rom-1}
\end{eqnarray}
where $\Delta t$ is the time step. 
The filtered convective term in~\eqref{eqn:l-rom-1} is defined as follows:
\begin{equation}    
	{\obu}_r^n(\bx,t) 
	\equiv \bU(\bx) 
	+ \sum_{j=1}^r \oa_j^n(t) \bphi_j(\bx)  \, .
	\label{eqn:l-rom-2}
\end{equation} 
The coefficients $\oa_j^n$ in~\eqref{eqn:l-rom-2} are defined as
\begin{equation}    
	\sum_{j=1}^r \oa_j^n(t) \bphi_j
	= \overline{ \sum_{j=1}^r a_j^n(t) \bphi_j } \, ,
	\label{eqn:l-rom-3}
\end{equation} 
where the filtering in~\eqref{eqn:l-rom-3} is effected by using the explicit ROM spatial filters defined in Section~\ref{sec:pod-projection} and Section~\ref{sec:differential-filter}.
We note that a forward Euler time discretization was used in \eqref{eqn:l-rom-2}, but other time discretizations are possible~\cite{ervin2012numerical}. 

\begin{remark}
	\label{remark:l-rom-1}
	We emphasize that, in~\eqref{eqn:l-rom-2}, only the unsteady (i.e., time-dependent) component of $\bu_r^n$ was filtered; the steady centering trajectory component was not filtered.
	This approach is consistent with the very purpose of Reg-ROMs, i.e., smoothing (regularizing) only the ROM terms that are the main contributors to the ROM's instability.
	Indeed, just as in the Fourier/spectral setting, the centering trajectory is generally the smoothest, most regular ROM component, whereas the smoothness of the velocity fluctuations decreases as their POD index increases.
	Thus, it is natural to expect that the velocity fluctuations and not the centering trajectory are generally responsible for the ROM's instability.
\end{remark}

The L-ROM~\eqref{eqn:l-rom-1} yields the following finite dimensional system:
For all $i = 1, \ldots, r$,
\begin{eqnarray}
	\dot{a}_i(t) 
	=  b_i
	+ \sum_{m=1}^{r} A_{im}a_m(t) 
	+ \sum_{m=1}^{r} \tA_{im}\oa_m(t) 
	+ \sum_{m=1}^r \sum_{n=1}^r B_{imn} \oa_m(t) a_n(t),
	\label{eqn:l-rom-4}
\end{eqnarray}
where
\begin{eqnarray}
	&& \hspace*{-1.3cm} 
	b_i 
	= -\bigl( {\bf U} \cdot \nabla {\bf U} , \bphi_i \bigr) 
	- \frac{2}{\mbox{Re}} \left( \frac{\nabla {\bf U} +\nabla {\bf U}^{\top}}{2} , \nabla \bphi_i \right), 
	\label{eqn:l-rom-5}\\
	&& \hspace*{-1.3cm} 
	A_{im} 
	= - \bigl( {\bf U} \cdot \nabla \bphi_m , \bphi_i \bigr) 
	- \frac{2}{\mbox{Re}} \left( \frac{\nabla \bphi_m 
	+ \nabla {\bphi_m}^{\top}}{2} , \nabla \bphi_i \right),
	\label{eqn:l-rom-6} \\
	&& \hspace*{-1.3cm} 
	\tA_{im} 
	= - \bigl(\bphi_m \cdot \nabla {{\bf U}} , \bphi_i \bigr),
	\label{eqn:l-rom-6} \\
	&& \hspace*{-1.3cm} 
	B_{imn} 
	= - \bigl( \bphi_m \cdot \nabla \bphi_n , \bphi_i \bigr) \, . 	
	\label{eqn:l-rom-8}
\end{eqnarray}

The Leray model was first used by Leray~\cite{leray1934sur} as a theoretical tool to prove local existence and uniqueness of weak solutions of the NSE.
The Leray model has been recently used as a numerical tool in the simulation of convection-dominated flows with standard (e.g., FE) numerical methods~\cite{layton2012approximate}.
To our knowledge, the Leray model was first used in a ROM setting in~\cite{sabetghadam2012alpha} for the numerical simulation of the 1D Kuramoto-Sivashinsky equation in a periodic setting.
In~\cite{sabetghadam2012alpha}, the spatial filter used was the DF. 
In this paper, we extend the L-ROM to the non-periodic 3D NSE.
Furthermore, we investigate both the DF and the Proj as spatial filters.

\subsection{The Evolve-Then-Filter ROM (EF-ROM)}
	\label{sec:ef-rom}

We propose a new Reg-ROM, the {\it evolve-then-filter ROM (EF-ROM)}: $\forall \, n = 0, \ldots, M$ and $\forall \, i = 1, \ldots, r,$
        
\begin{eqnarray}
        &&	\left(
                \frac{\bw_{r}^{n+1} - \bu_{r}^{n}}{\Delta t} , \bphi_{i}
            \right)
            + \frac{2}{Re} \, \biggl(
               \mathbbm{D}(\bu_{r}^{n}) ,
                \nabla \bphi_{i}
            \biggr)
            + \biggl(
                (\bu_{r}^{n} \cdot \nabla) \bu_{r}^{n} ,
                \bphi_{i}
            \biggr)
            = 0 \, ,
            \label{eqn:ef-rom-1}  \\
            && \bu_{r}^{n+1}
            = \overline{\bw_{r}^{n+1}} \, ,
            \label{eqn:ef-rom-2}
\end{eqnarray}
where $\Delta t$ is the time step. 
The ``evolve" step in the EF-ROM (i.e., equation~\eqref{eqn:ef-rom-1}) is just one step of the time discretization of the standard G-ROM~\eqref{eqn:g-rom-2}. 
The ``filter" step in the EF-ROM (i.e., equation~\eqref{eqn:ef-rom-2}) consists of filtering of the intermediate solution obtained in the ``evolve" step. 
This filtering in~\eqref{eqn:ef-rom-2} is defined as in~\eqref{eqn:l-rom-2}--\eqref{eqn:l-rom-3}, i.e., its unsteady component is filtered, but its steady component (i.e., the centering trajectory) is not filtered.
This strategy is consistent with the Reg-ROM strategy (see Remark~\ref{remark:l-rom-1}).
As mentioned in Section~\ref{sec:l-rom}, a forward Euler time discretization was used in \eqref{eqn:ef-rom-1}, but other time discretizations are possible~\cite{ervin2012numerical}. 

The evolve-then-filter model has been used as a numerical tool in the simulation of convection-dominated flows with standard (e.g., FE or spectral element) numerical methods~\cite{layton2012approximate}.
We emphasize, however, that, to our knowledge, the evolve-then-filter model has not been used in a ROM setting.

\section{Numerical Tests}
    \label{sec:numerical-tests}

\noindent {\bf Goals} \ 
This section has two goals: 
The first goal is to compare the EF-ROM~\eqref{eqn:ef-rom-1}--\eqref{eqn:ef-rom-2} with the L-ROM~\eqref{eqn:l-rom-1}. 
The second goal is to investigate the effect of explicit ROM spatial filtering (the Proj and the DF) on the L-ROM and EF-ROM. 
Numerical results are presented for a 3D flow past a cylinder at Reynolds number $Re=1000$.
We also include results for the G-ROM~\eqref{eqn:g-rom-2}.
A successful Reg-ROM should at least perform better than the G-ROM.
Finally, a DNS projection of the evolution of the POD modes serves as benchmark for our numerical 
simulations. \\[-0.3cm]

\noindent {\bf Models} \ 
    For each of the two Reg-ROMs (the L-ROM and the EF-ROM), two distinct
    filtering strategies are considered: the Proj and the DF. 
    Thus, four Reg-ROM/filter combinations are investigated. For
    clarity, the following notation is used for the resulting Reg-ROM/filter
    combinations: L-ROM-DF for the L-ROM with the DF,
    L-ROM-Proj for the L-ROM with the Proj, EF-ROM-DF for the EF-ROM
    with the DF, and EF-ROM-Proj for the EF-ROM with the
    Proj. \\[-0.3cm]

\noindent {\bf Criteria} \ 
The qualitative behavior of the ROMs is judged according to the following six criteria~\cite{wang2012proper}: 
(i) the kinetic energy spectrum;
(ii) the mean velocity; 
(iii) the Reynolds stresses;
(iv) the root mean square (rms) values of the velocity fluctuations;
(v) the time evolution of the POD coefficients; and
(vi) the Strouhal number.
The first four criteria are statistics that measure the temporal and spatial average behavior of the ROMs, whereas the fifth criterion measures the instantaneous behavior of the ROMs.
We also include a computational efficiency assessment for all the ROMs.
Next, we briefly present these criteria.
To this end, we first describe the averaging operator, which is used in all the statistics.
In this paper, we use the averaging operator $\langle \cdot \rangle = \langle \cdot \rangle_{tyz}$, which consists of averaging in time (over the time interval $[0,300]$) and in the $y$- and $z$- directions. 
Specifically, to compute $\langle q \rangle_{tyz}$ for a given quantity $q$, for each fixed point $x$, we have
\begin{equation}
	\langle q \rangle_{tyz}(x) 
	= \frac{1}{T \, L_y \, L_z} \sum_{t, y, z} q(x, y, z, t)  \, ,
\end{equation}
where $T$ is the total time length (i.e., $T=300$), $L_y$ is the dimension of the computational domain in the $y$-direction, and $L_z$ is the dimension of the computational domain in the $z$-direction. 
Since  the topology of the velocity field is markedly different in the $x$-, $y$-, and $z$-directions, one could also consider spatial averaging in the $xz$-direction ($\left< \cdot \right> = \left< \cdot \right>_{txz}$).
We note that, since the numerical results with spatial averaging in the $xz$-direction are qualitatively similar to those with spatial averaging in the $yz$-direction, they are not included in the paper. \\[-0.4cm]

\noindent {\it Energy Spectrum}: \ 
All energy spectra are calculated from the average kinetic energy of the nodes in the cube with side $0.1$ centered at the probe $(0.9992,0.3575,1.0625)$. \\[-0.4cm]

\noindent {\it Mean Velocity Components}: \ 
The following mean velocity components are plotted: $\langle u \rangle$ (the mean streamwise velocity), $\langle v \rangle$ (the mean normal velocity), and $\langle w \rangle$ (the mean spanwise velocity). \\[-0.4cm]

\noindent {\it Reynolds Stresses}: \ 
The following Reynolds stresses are plotted: $\langle u  - \langle u \rangle , v - \langle v \rangle\rangle$ (the xy-component of the Reynolds stress), $\langle u  - \langle u \rangle , w - \langle w \rangle\rangle$ (the xz-component of the Reynolds stress), and $\langle v  - \langle v \rangle , w - \langle w \rangle\rangle$ (the yz-component of the Reynolds stress). \\[-0.4cm]

\noindent {\it RMS Values of Velocity Fluctuations}: \ 
The following rms values of velocity fluctuations are plotted:  $\langle u \rangle_{rms} = \langle u  - \langle u \rangle , u - \langle u \rangle\rangle$ (the rms of the streamwise velocity fluctuations), $\langle v \rangle_{rms} = \langle v  - \langle v \rangle , v - \langle v \rangle\rangle$ (the rms of the normal velocity fluctuations), and $\langle w \rangle_{rms} = \langle w  - \langle w \rangle , w - \langle w \rangle\rangle$ (the rms of the spanwise velocity fluctuations). \\[-0.4cm]

\noindent {\it Strouhal Number}: \ 
The Strouhal number ($St$) is computed as~\cite{akhtar2008parallel,akhtar2009stability}
\begin{equation}
	St 
	= \frac{f \, D}{U_{\infty}} \, ,
\end{equation}
where $f$ is the shedding frequency, $U_{\infty}$ is the streamwise velocity (i.e., the velocity at the inlet), and $D$ is the diameter of the cylinder.
Since we have limited the scope of this paper to just velocity ROMs, we do not consider a ROM representation of the pressure field. 
Hence we do not calculate the lift and drag coefficient spectra. 
Thus, we compute the shedding frequency $f$ as the frequency corresponding to the first spike in the energy spectrum (see discussion on page 410 in~\cite{kravchenko2000numerical}). \\[0.3cm]

\noindent {\bf Numerical Methods} \
We investigate all ROMs in the numerical simulation of 3D flow past a circular cylinder at $\mbox{Re} = 1000$. 
The Reynolds number is computed using the diameter ($D$) of the cylinder as the length scale and the freestream velocity ($U_\infty$) as the velocity scale. 
The cylinder is parallel to the $z$-axis and the free-stream flow is in the positive $x$-direction.
In this section,  $u$ denotes the streamwise velocity component (associated with the $x$-axis), $v$ denotes the normal velocity component (associated with the $y$-axis), and $w$ denotes the spanwise velocity component (associated with the $z$-axis).
A parallel CFD solver is employed on the time interval $[0,300]$ to generate the DNS data. 
Details on the numerical discretization are presented in~\cite{akhtar2008parallel,akhtar2015using,akhtar2009stability}.

We obtain the POD basis by collecting $1000$ snapshots of the velocity field ($u,v,w$) over the time interval $[0,75]$ and applying the method of snapshots developed in \cite{Sir87abc}. 
These POD modes are then interpolated onto a structured quadratic FE triangulation 
with nodes coinciding with the nodes used in the original DNS finite difference discretization. 
The first $r=6$ POD modes capture 84\% of the energy of the velocity fluctuations.
These modes are used in all ROMs.
For all the ROMs, the time discretization employs the explicit Euler method with $\Delta t= 7.5\times10^{-4}$ and the spatial discretization uses piecewise quadratic Lagrange FEs. 
All the ROMs are investigated on the time interval $[0,300]$.
\\[-0.3cm]

\noindent {\bf Parameters} \
The Reg-ROM results are plotted for the optimal values of $\delta$ (for L-ROM-DF and EF-ROM-DF) and $r_1$ (for L-ROM-Proj and EF-ROM-Proj). 
These optimal $\delta$ and $r_1$ values are determined by requiring that the Reg-ROMs have the time average of the $L^2$-norm of the solution as close as possible to that of the DNS. 
To find the optimal $\delta$ value for L-ROM-DF and EF-ROM-DF, the time averages of the $L^2$-norms of the solutions of these Reg-ROMs are plotted for $101$ $\delta$ values in the interval $[0,75]$. 
The intersection between the Reg-ROM and DNS curves yields the optimal $\delta$ value. 
To find the optimal $r_1$ value for the L-ROM-Proj and EF-ROM-Proj, a similar approach is used. 
This time, however, since $r_1$ takes integer values, only a relatively low number of values is used in the optimization procedure.
This approach yields the following values for the optimal parameters:  $r_1 = 1$ for L-ROM-Proj,  $\delta = 0.247$ for L-ROM-DF, $r_1 = 4$ for EF-ROM-Proj, and $\delta = 0.001367$ for EF-ROM-DF.
We emphasize that these parameter values are optimal only on the short time interval tested (i.e., $[0,75]$), and they might actually be non-optimal on the entire time interval $[0,300]$ on which the Reg-ROMs are tested.
Thus, this heuristic procedure ensures some fairness in the numerical comparison of the four Reg-ROMs.

\subsection{Numerical Results}
	\label{sec:numerical-results}

\begin{figure}
	\centering
	\caption{
		Snapshots of horizontal velocity at $t = 142.5 s$ for: 
		the DNS (top, left);
		the G-ROM (top, right);
		the L-ROM-Proj (middle, left); 
		the L-ROM-DF (middle, right);
		the new EF-ROM-Proj (bottom, left); and
		the new EF-ROM-DF (bottom, right).
		Seven isosurfaces are plotted.
		\label{fig:snapshots}
	}
	\begin{minipage}[h]{0.45\linewidth} \includegraphics[width=0.9\textwidth]{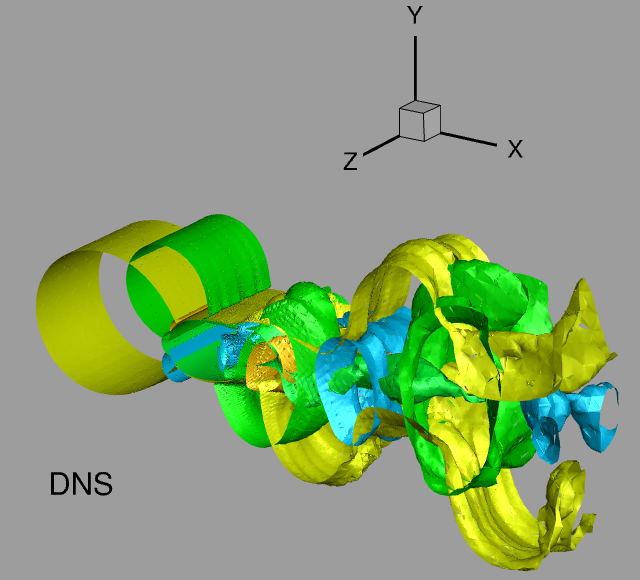}\end{minipage}
	\begin{minipage}[h]{0.45\linewidth} \includegraphics[width=0.9\textwidth]{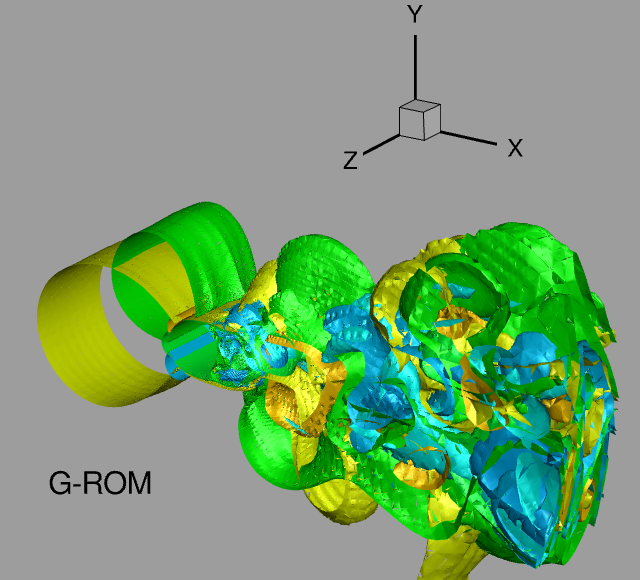}\end{minipage}\\[0.8cm]
	\begin{minipage}[h]{0.45\linewidth} \includegraphics[width=0.9\textwidth]{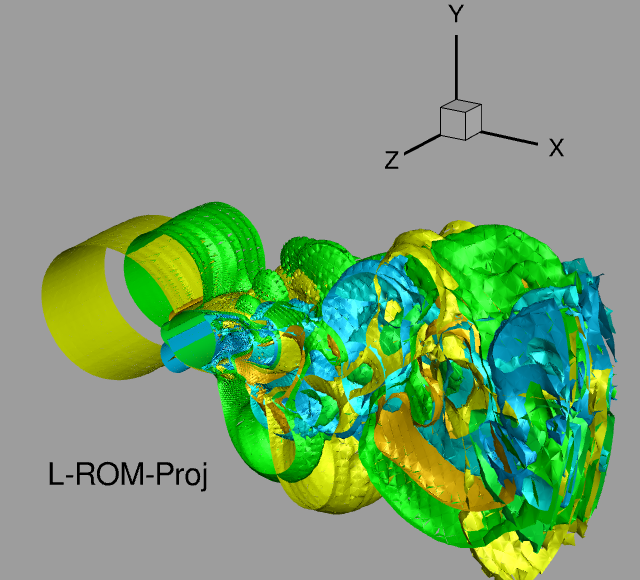}\end{minipage}
	\begin{minipage}[h]{0.45\linewidth} \includegraphics[width=0.9\textwidth]{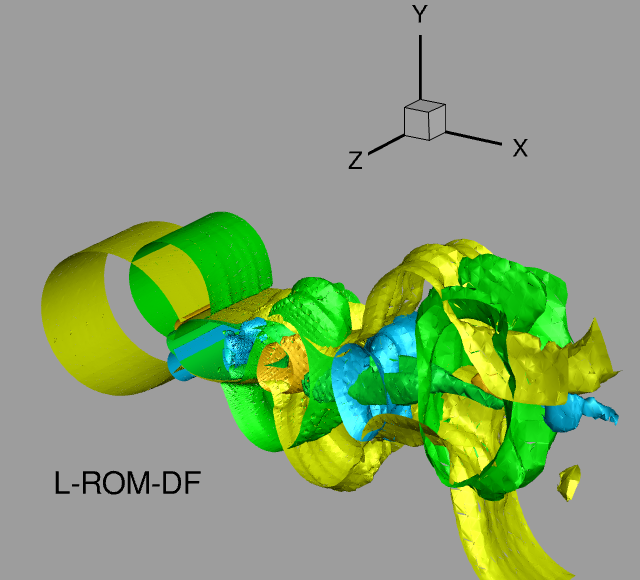}\end{minipage}\\[0.8cm]
	\begin{minipage}[h]{0.45\linewidth} \includegraphics[width=0.9\textwidth]{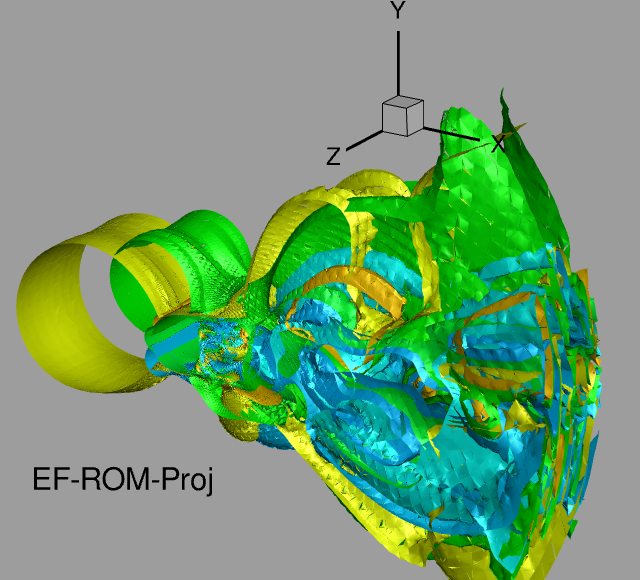}\end{minipage}
	\begin{minipage}[h]{0.45\linewidth} \includegraphics[width=0.9\textwidth]{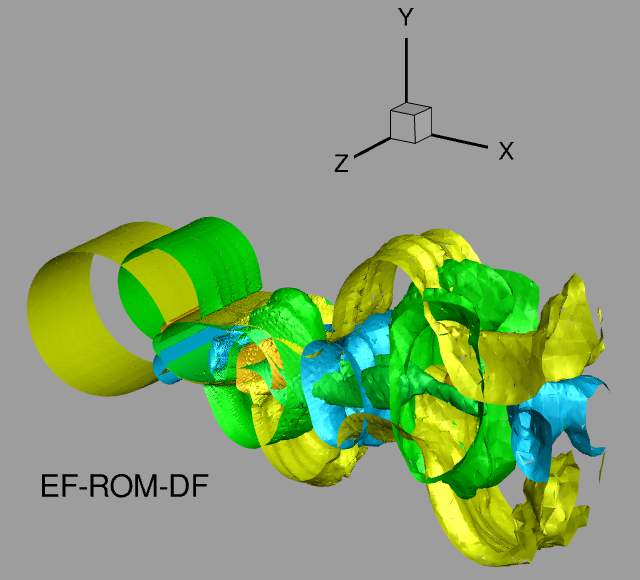}\end{minipage}
\end{figure}

Before presenting the quantitative comparison of the ROMs, we give a flavor of the topology of the resulting flow fields. 
In Fig.~\ref{fig:snapshots}, we plot seven isosurfaces of the velocity snapshots at $t=142.5\,s$ for DNS, G-ROM, L-ROM-Proj, L-ROM-DF, EF-ROM-Proj, and EF-ROM-DF.
Taking the DNS results as a benchmark, the G-ROM, EF-ROM-Proj L-ROM-Proj yield inaccurate results, in the form of unphysical structures.
The L-ROM-DF yields more accurate results.
The most accurate results, however, are produced by the EF-ROM-DF.
Due to space limitations, only one time instance snapshot is shown for the ROMs.
The general behavior over the entire time interval is similar.

In Fig.~\ref{fig:spectrum}, we plot the energy spectra of the four ROMs and, for comparison purposes, of the G-ROM.
The five energy spectra are compared with the DNS energy spectrum.
The energy spectra of the G-ROM and EF-ROM-Proj overestimate the energy spectrum of the DNS.
The energy spectra of the L-ROM-Proj and L-ROM-DF, on the other hand, underestimate the energy spectrum of the DNS.
The most accurate energy spectrum (i.e., the closest to the DNS energy spectrum) is clearly that of the EF-ROM-DF.

In Fig.~\ref{fig:mean}, we plot the mean velocity components for different values of $x$.
Figure~\ref{fig:mean} yields the following conclusions:
First, the mean streamwise velocity is computed accurately by all ROMs.
Second, the G-ROM and EF-ROM-Proj yield inaccurate results for the mean spanwise velocity; all the other ROMs perform significantly better.
Third, the EF-ROM-Proj yields inaccurate results for the mean normal velocity; all the other ROMs perform significantly better.

In Fig.~\ref{fig:stress}, we plot the Reynolds stresses for different values of $x$.
Figure~\ref{fig:stress} yields the following conclusions:
First, the EF-ROM-Proj and G-ROM Reynolds stresses are consistently the most inaccurate (i.e., the farthest
from the DNS Reynolds stresses).
Second, the other ROMs have all similar behaviors.
Indeed, different ROMs might outperform the others over different spatial regions, but there is no clear ``winner" over the entire spatial interval for any of the Reynolds stresses.

In Fig.~\ref{fig:rms}, we plot the rms values for different values of $x$.
Figure~\ref{fig:rms} yields the following conclusions:
Similar to the Reynolds stresses case, the EF-ROM-Proj and G-ROM rms values of the velocity fluctuations are consistently the most inaccurate (i.e., the farthest from the DNS rms values).
The rms plots corresponding to the other ROMs, however, display a consistent ordering  this time.
Indeed, when the $\left< v \right>_{rms}$ and $\left< w \right>_{rms}$ plots are considered, the EF-ROM-DF is slightly better than the L-ROM-DF.
Furthermore, both are better than L-ROM-Proj. 
When the $\left< u \right>_{rms}$ plot is considered, the EF-ROM-DF, L-ROM-DF, and L-ROM-Proj perform similarly.

The time evolutions of the POD basis coefficients $a_2(\cdot)$ and $a_6(\cdot)$ on the entire time interval $[0, 300]$ are shown in  Fig.~\ref{fig:a2-a6}. 
We note that the other POD coefficients have a similar behavior. 
Thus, for clarity of exposition, we include only $a_2(\cdot)$ and $a_6(\cdot)$. 
The EF-ROM-Proj and G-ROM's time evolutions of $a_2$ and $a_6$ are clearly inaccurate. 
Indeed, the magnitudes $a_2$ and $a_6$ are several times larger than that of the DNS projection. 
The L-ROM-Proj and, to a smaller extent, L-ROM-DF also yield inaccurate time evolutions of $a_2$ and $a_6$.
This time, however, these ROMs' POD coefficients seem to decrease as time increases.
The EF-ROM-DF clearly yields the most accurate POD coefficients.

The Strouhal numbers predicted by the ROMs and the DNS are listed in Table~\ref{table:strouhal}.
It is clear that the EF-ROM-DF yields the most accurate prediction of the Strouhal number, whereas the EF-ROM-Proj yields the most inaccurate prediction.
Also note that the G-ROM fails to predict a clear shedding frequency and, thus, cannot yield a Strouhal number.

\begin{table}[h]
  \begin{center}
  		\caption{Strouhal numbers predicted by ROMs.}
		\label{table:strouhal}
	\begin{tabular}{ccccccc}
		\hline
		{} & {\scriptsize DNS} & {\scriptsize G-ROM}&{\scriptsize L-ROM-Proj} & {\scriptsize L-ROM-DF} & {\scriptsize EF-ROM-Proj} & {\scriptsize EF-ROM-DF} \\
		\hline
		$St$ & 0.2083 & -- & 0.1855 & 0.1855 & 0.2474 & 0.1986 \\
		\hline
	\end{tabular}
  \end{center}
\end{table}

A natural question, however, is whether the Reg-ROMs that we investigated are computationally efficient, which is one of the main requirements for any successful ROM.
The CPU times of all four Reg-ROM/filter combinations and G-ROM are listed in Table \ref{cpu-times}.
The CPU time of the Reg-ROMs is on the same order as the CPU time of the G-ROM.
In~\cite{wang2012proper}, we showed that the CPU time of the G-ROM is orders of magnitude lower than the CPU time of the DNS.
Thus, we conclude that the Reg-ROMs' computational efficiency is extremely high, similar to that of the G-ROM.

\begin{table}[h]
  \begin{center}
  		\caption{CPU times (in seconds) of ROMs.}
		\label{cpu-times}
	\begin{tabular}{ccccc}
		\hline
		{\scriptsize G-ROM}&{\scriptsize L-ROM-Proj} & {\scriptsize L-ROM-DF} & {\scriptsize EF-ROM-Proj} & {\scriptsize EF-ROM-DF} \\
		\hline
		92 & 103 & 104 & 114& 101 \\
		\hline
	\end{tabular}
  \end{center}
\end{table}

\subsection{Summary and Discussion}
	\label{sec:summary-discussion}

\begin{figure}
	\centering
	\caption{
		Kinetic energy spectrum of the DNS (blue) and the ROMs (red):
		(a) the G-ROM;
		(b) the L-ROM-Proj; 
		(c) the L-ROM-DF;
		(d) the EF-ROM-Proj; and
		(e) the EF-ROM-DF.
		\label{fig:spectrum}
	}
	\begin{minipage}[h]{0.03\linewidth} (a) \end{minipage}
	\begin{minipage}[h]{0.8\linewidth} \includegraphics[width=0.9\textwidth]{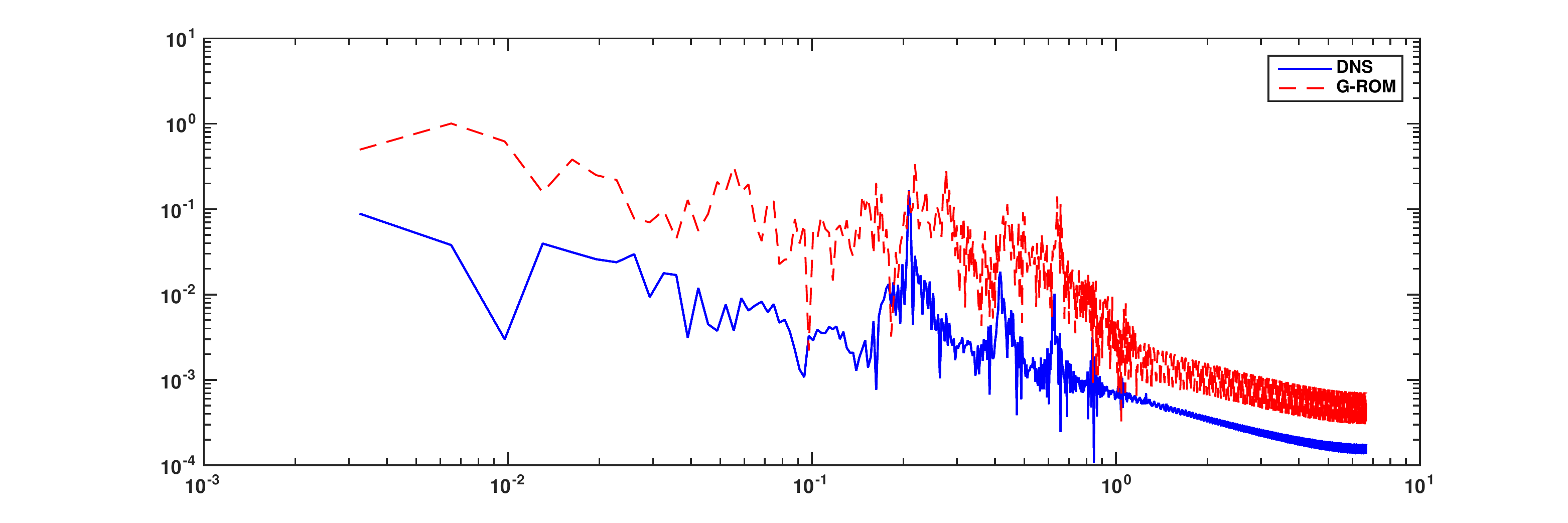}\end{minipage}\\
	\begin{minipage}[h]{0.03\linewidth} (b) \end{minipage}
	\begin{minipage}[h]{0.8\linewidth} \includegraphics[width=0.9\textwidth]{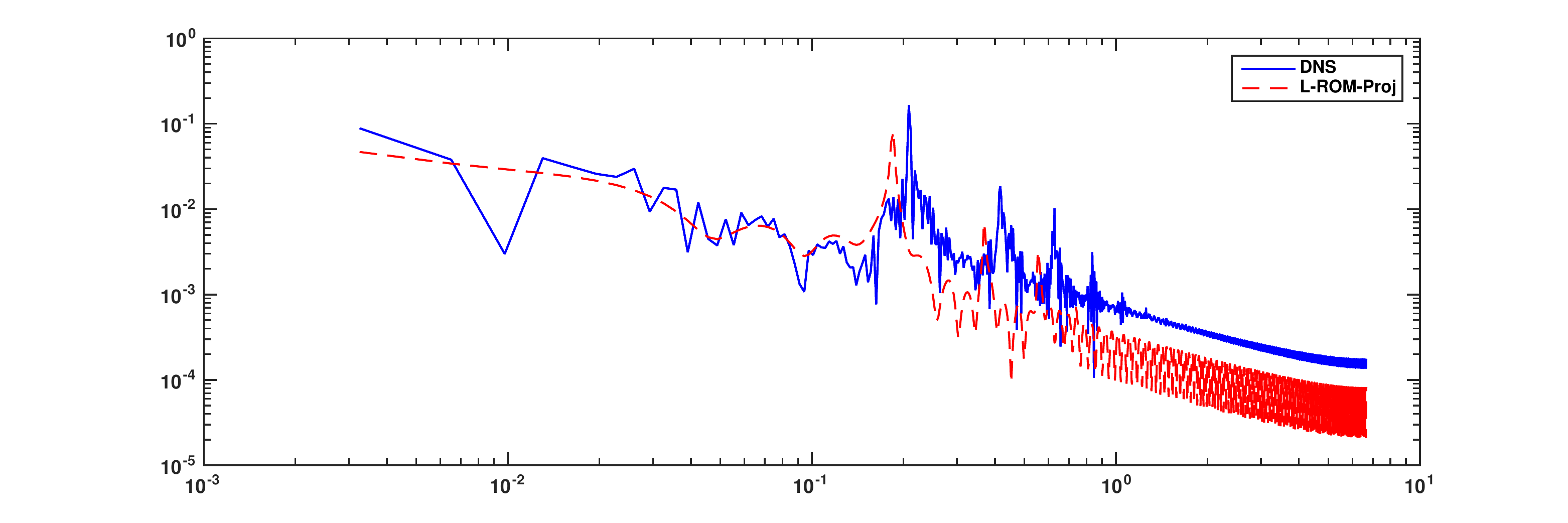}\end{minipage}\\
	\begin{minipage}[h]{0.03\linewidth} (c) \end{minipage}
	\begin{minipage}[h]{0.8\linewidth} \includegraphics[width=0.9\textwidth]{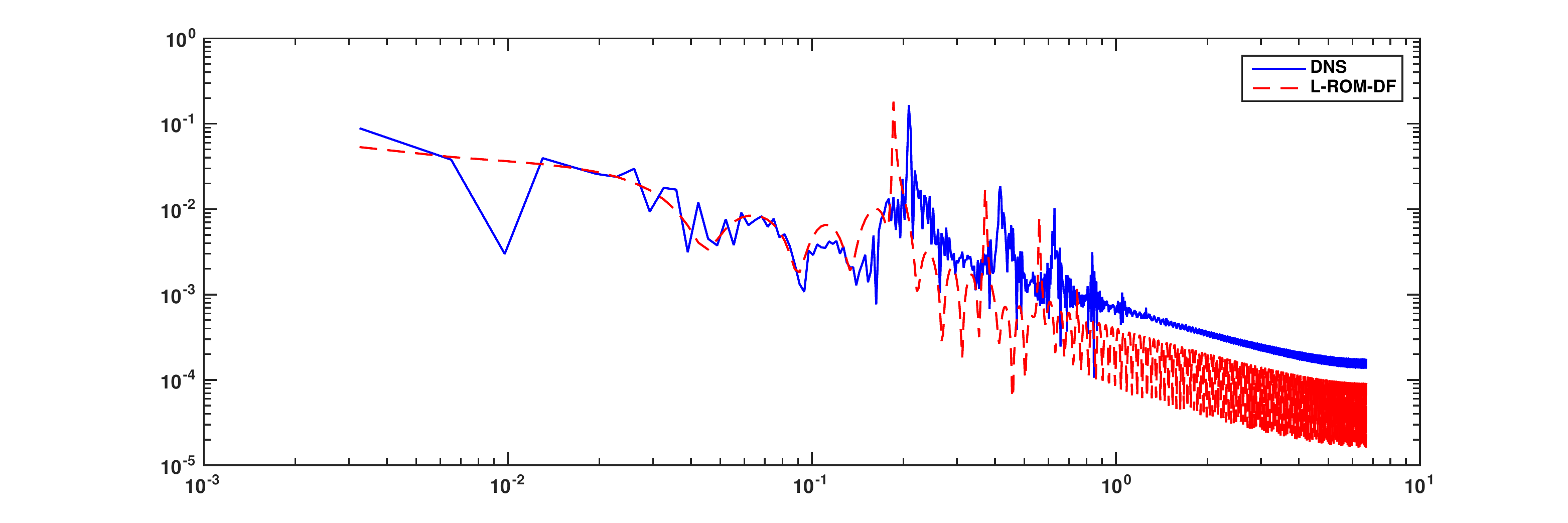}\end{minipage}\\
	\begin{minipage}[h]{0.03\linewidth} (d) \end{minipage}
	\begin{minipage}[h]{0.8\linewidth} \includegraphics[width=0.9\textwidth]{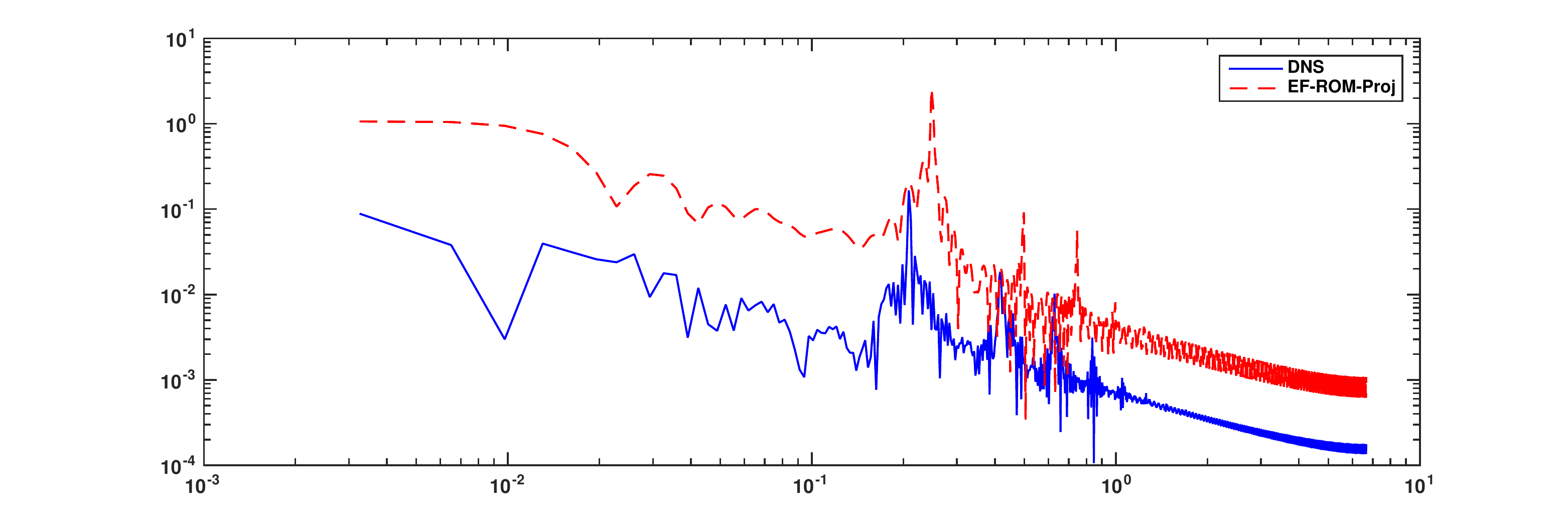}\end{minipage}\\
	\begin{minipage}[h]{0.03\linewidth} (e) \end{minipage}
	\begin{minipage}[h]{0.8\linewidth} \includegraphics[width=0.9\textwidth]{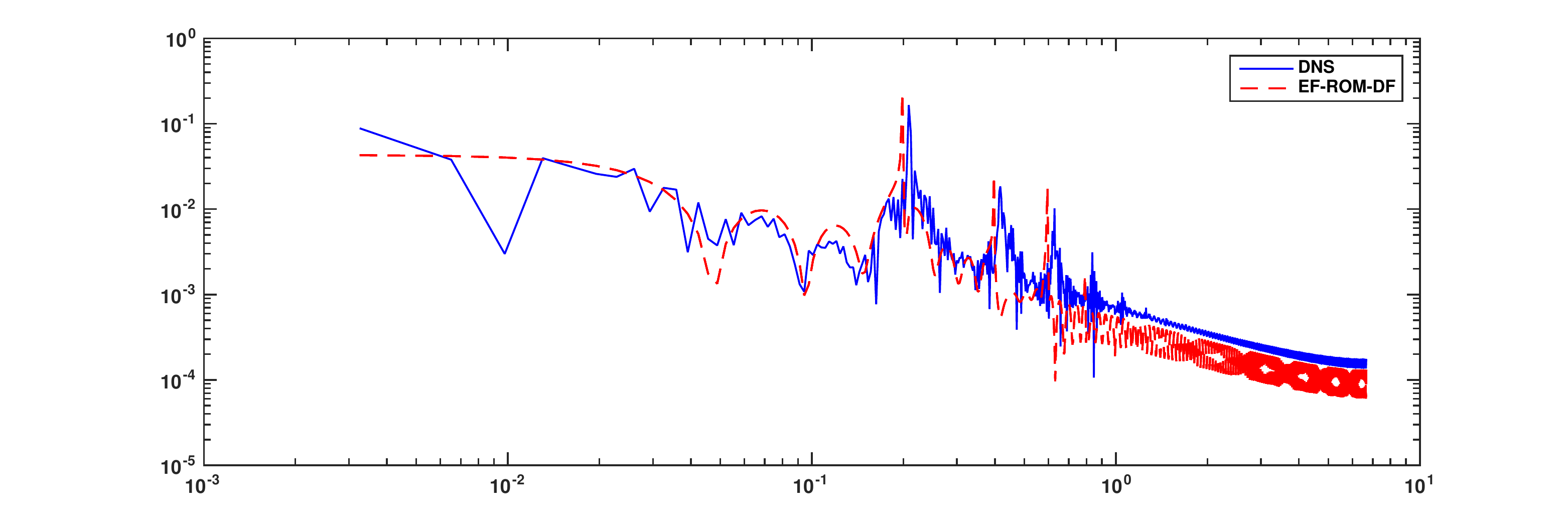}\end{minipage}\\
\end{figure}

\begin{figure}
	\centering
	\caption{
		Mean velocity components of DNS and ROMs:
		(a) $\left< u \right>$ (the mean streamwise velocity),
		(b) $\left< v \right>$ (the mean spanwise velocity), and
		(c) $\left< w \right>$ (the mean normal velocity).
		\label{fig:mean}
	}
	\begin{minipage}[h]{0.03\linewidth} (a) \end{minipage}
	\begin{minipage}[h]{0.7\linewidth} \includegraphics[width=0.9\textwidth]{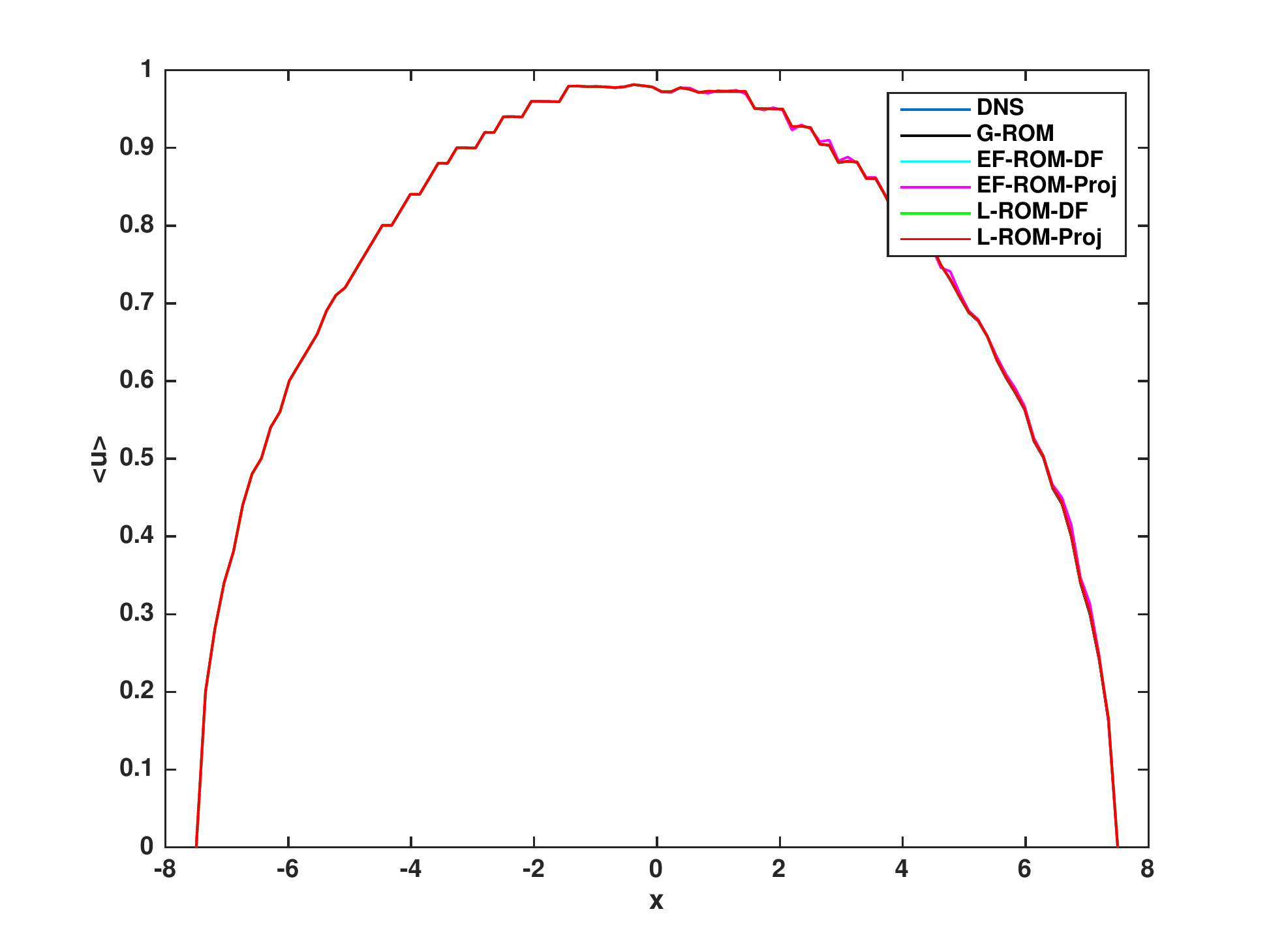}\end{minipage}\\
	\begin{minipage}[h]{0.03\linewidth} (b) \end{minipage}
	\begin{minipage}[h]{0.7\linewidth} \includegraphics[width=0.9\textwidth]{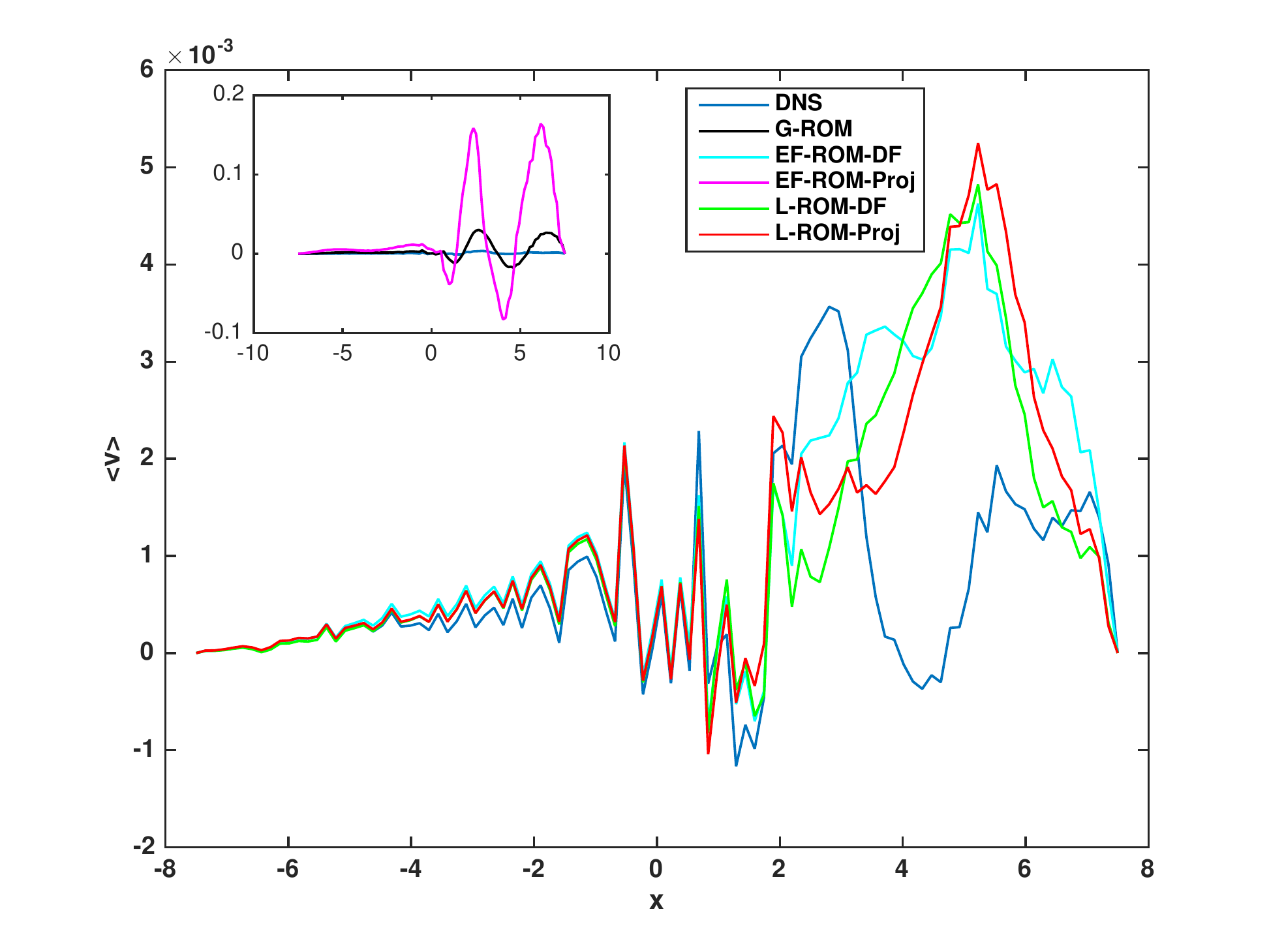}\end{minipage}\\
	\begin{minipage}[h]{0.03\linewidth} (c) \end{minipage}
	\begin{minipage}[h]{0.7\linewidth} \includegraphics[width=0.9\textwidth]{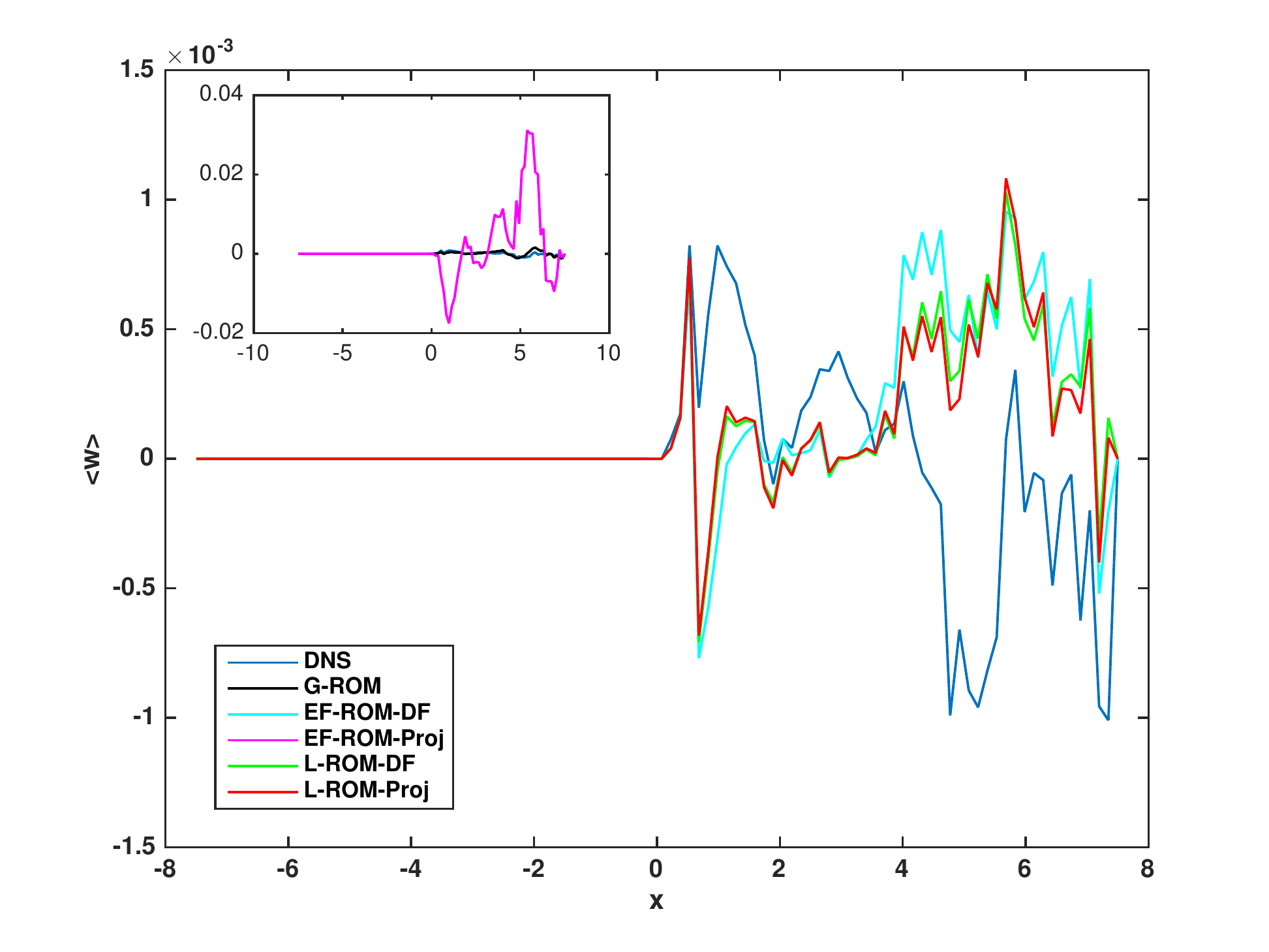}\end{minipage}\\
\end{figure}

\begin{figure}
	\centering
	\caption{
		Reynolds stresses of DNS and ROMs:
		(a) $\left< u - \left< u \right> , v - \left< v \right>\right>$ (the $xy$-component of the Reynolds stress),
		(b) $\left< u - \left< u \right> , w - \left< w \right>\right>$ (the $xz$-component of the Reynolds stress), and
		(c) $\left< v - \left< v \right> , w - \left< w \right>\right>$ (the $yz$-component of the Reynolds stress).
		\label{fig:stress}
	}
	\begin{minipage}[h]{0.03\linewidth} (a) \end{minipage}
	\begin{minipage}[h]{0.7\linewidth} \includegraphics[width=0.9\textwidth]{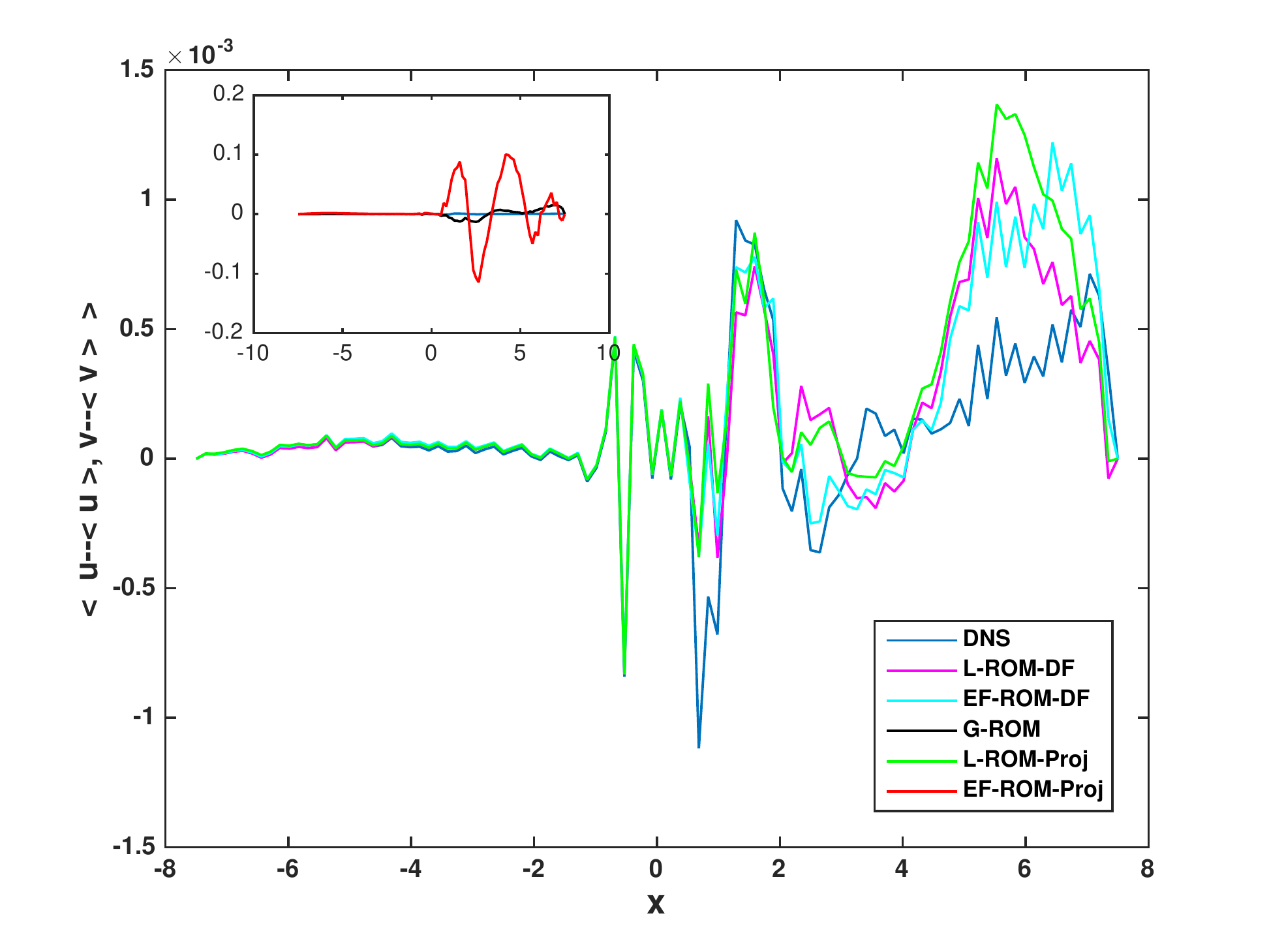}\end{minipage}\\
	\begin{minipage}[h]{0.03\linewidth} (b) \end{minipage}
	\begin{minipage}[h]{0.7\linewidth} \includegraphics[width=0.9\textwidth]{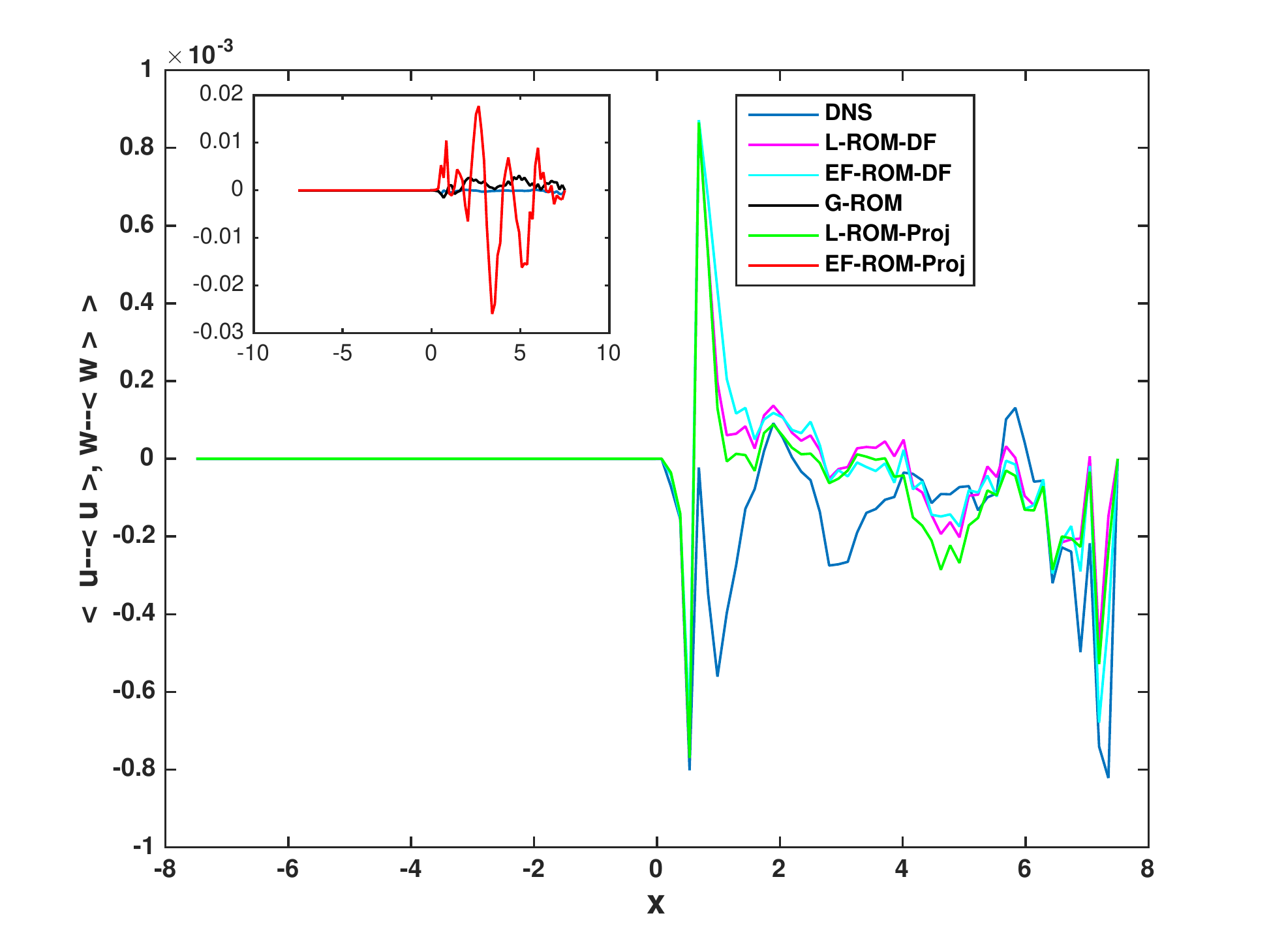}\end{minipage}\\
	\begin{minipage}[h]{0.03\linewidth} (c) \end{minipage}
	\begin{minipage}[h]{0.7\linewidth} \includegraphics[width=0.9\textwidth]{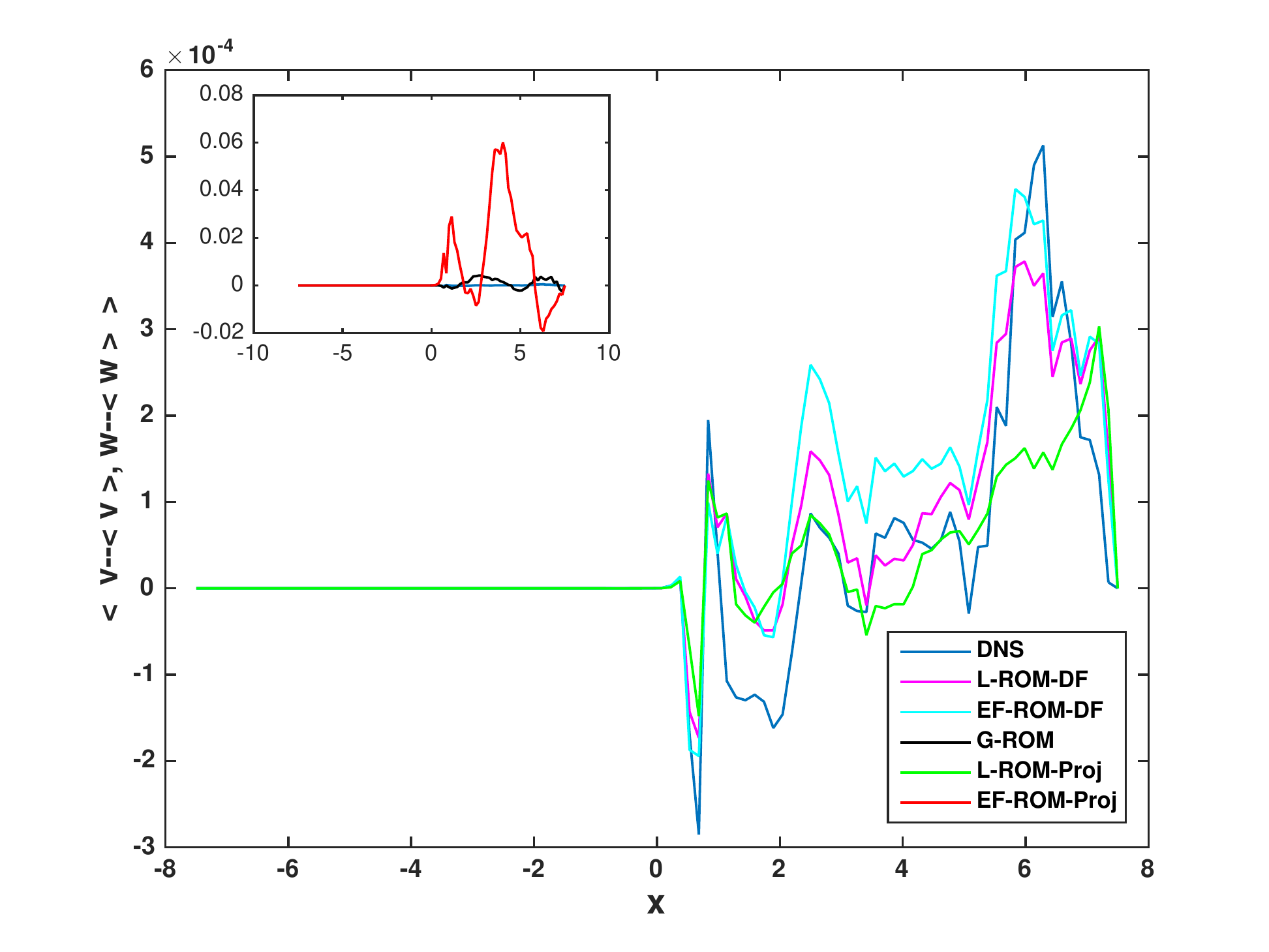}\end{minipage}\\
\end{figure}

\begin{figure}
	\centering
	\caption{
		Rms values of the velocity fluctuations of DNS and ROMs:
		(a) $\left< u \right>_{rms} = \left< u - \left< u \right> , u - \left< u \right> \right>$ (the rms value of the streamwise velocity fluctuations),
		(b) $\left< v \right>_{rms} = \left< v - \left< v \right> , v - \left< v \right> \right>$ (the rms value of the spanwise velocity fluctuations), and
		(c) $\left< w \right>_{rms} = \left< w - \left< w \right> , w - \left< w \right> \right>$ (the rms value of the normal velocity fluctuations).
		\label{fig:rms}
	}
	\begin{minipage}[h]{0.03\linewidth} (a) \end{minipage}
	\begin{minipage}[h]{0.7\linewidth} \includegraphics[width=0.9\textwidth]{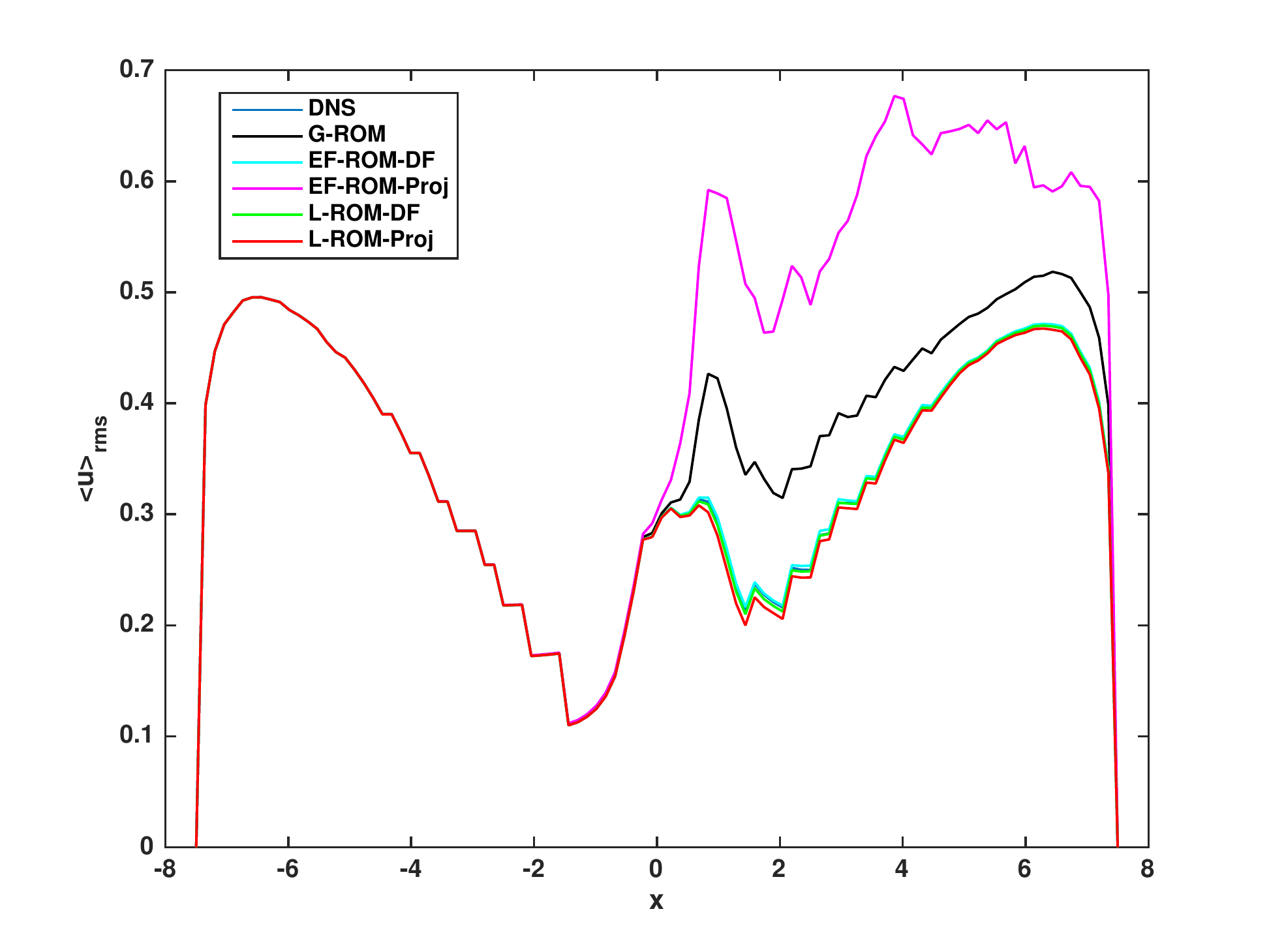}\end{minipage}\\
	\begin{minipage}[h]{0.03\linewidth} (b) \end{minipage}
	\begin{minipage}[h]{0.7\linewidth} \includegraphics[width=0.9\textwidth]{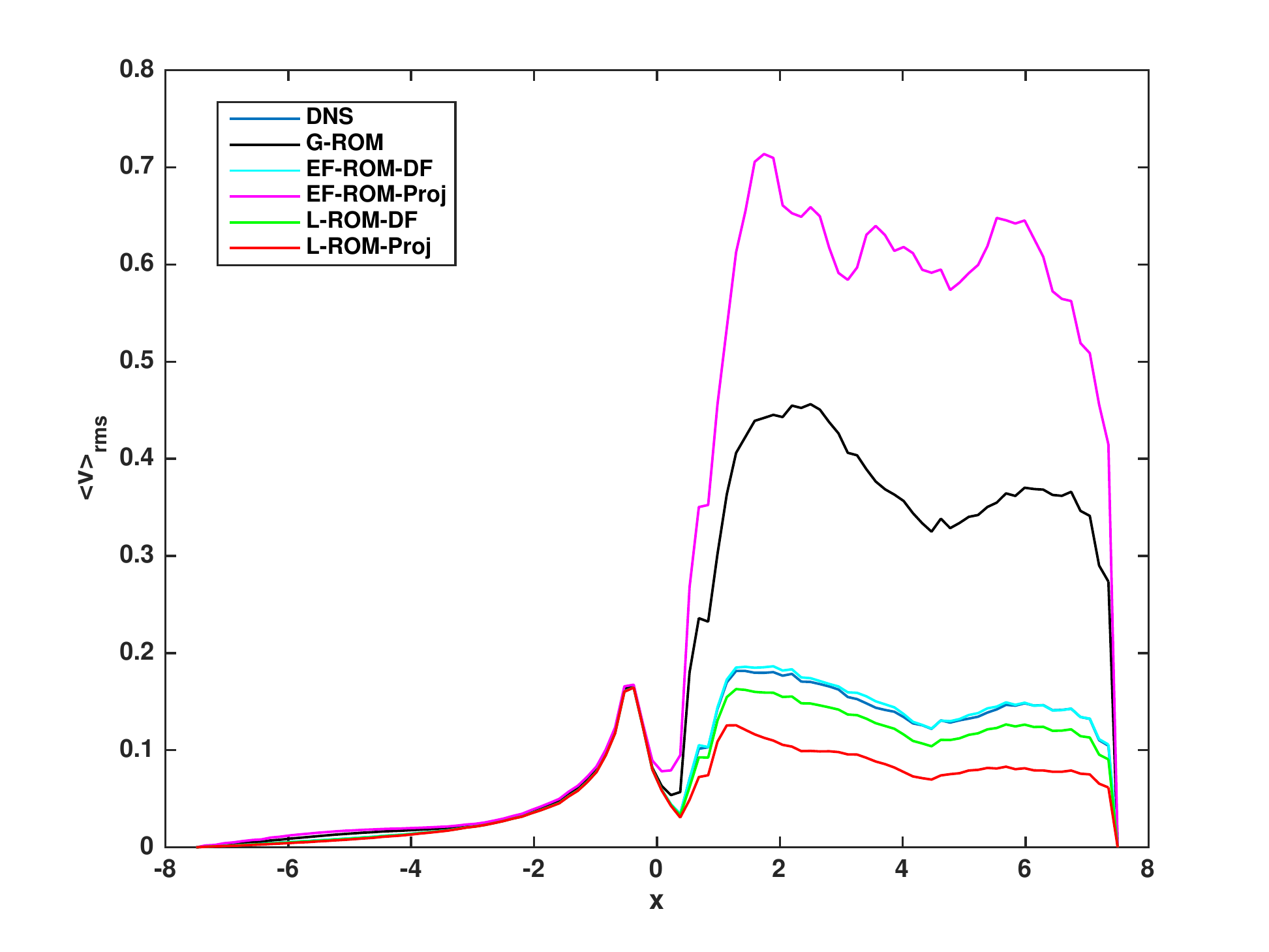}\end{minipage}\\
	\begin{minipage}[h]{0.03\linewidth} (c) \end{minipage}
	\begin{minipage}[h]{0.7\linewidth} \includegraphics[width=0.9\textwidth]{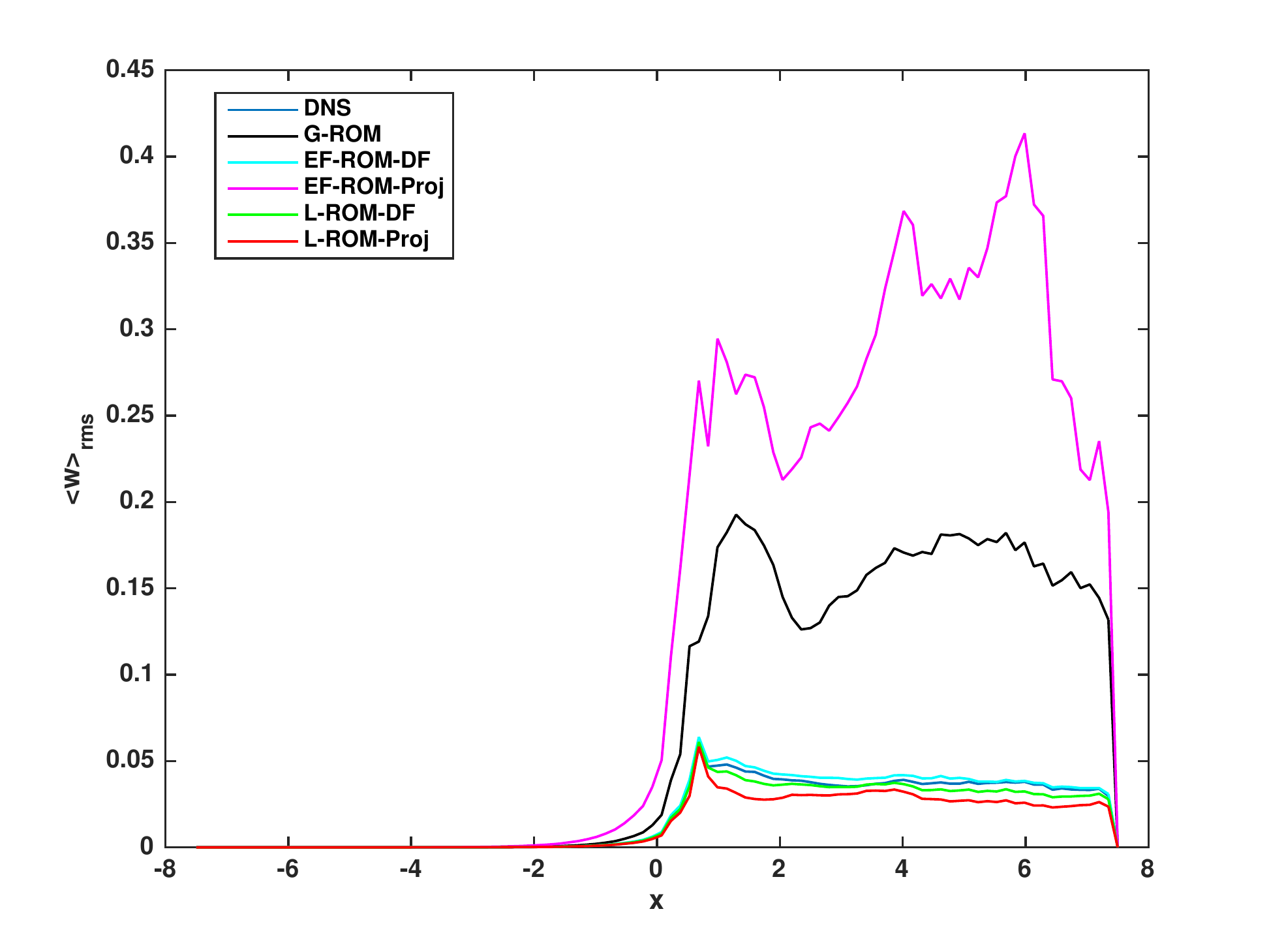}\end{minipage}\\
\end{figure}

\begin{figure}
	\centering
	\caption{
		Time evolution of the POD basis coefficients $a_2$ (left column) and $a_6$ (right column) of the DNS (blue) and ROMs (red):
		the G-ROM (first row);
		the L-ROM-Proj (second row); 
		the L-ROM-DF (third row);
		the new EF-ROM-Proj (fourth row); and
		the new EF-ROM-DF (fifth row).
		\label{fig:a2-a6}
	}
	\begin{minipage}[h]{0.4\linewidth} \includegraphics[width=0.9\textwidth]{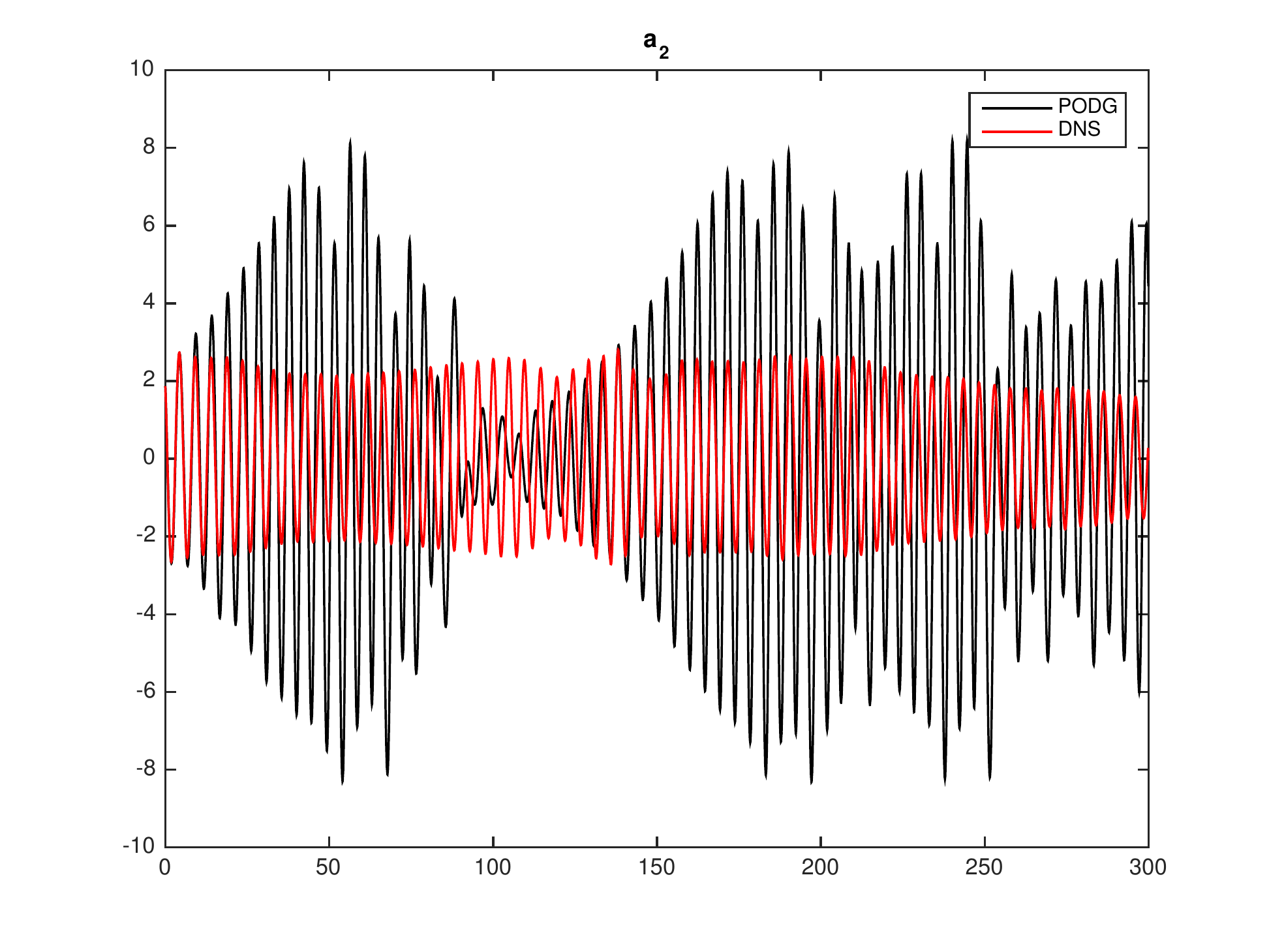}\end{minipage}
	\begin{minipage}[h]{0.4\linewidth} \includegraphics[width=0.9\textwidth]{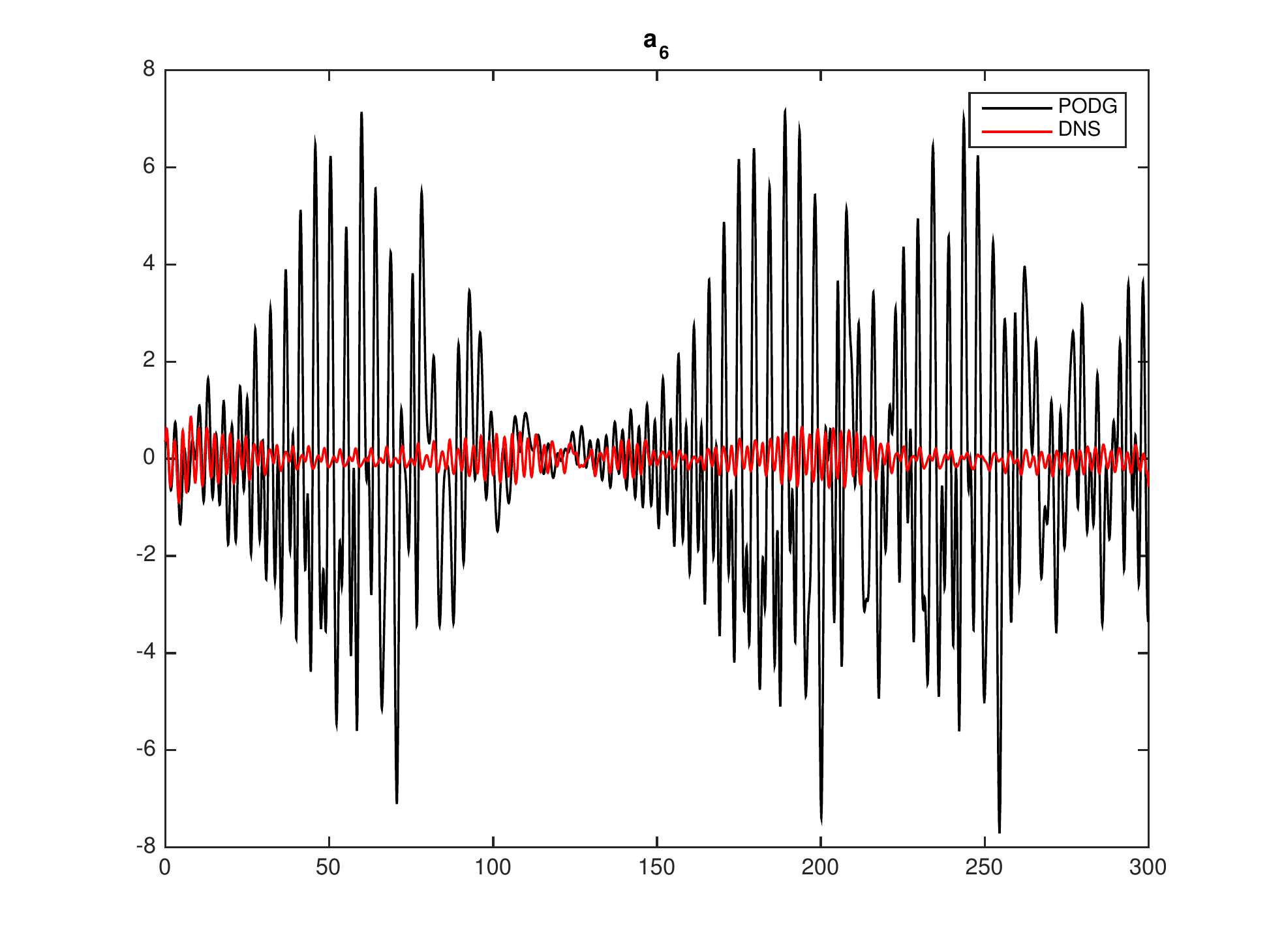}\end{minipage}\\
	\begin{minipage}[h]{0.4\linewidth} \includegraphics[width=0.9\textwidth]{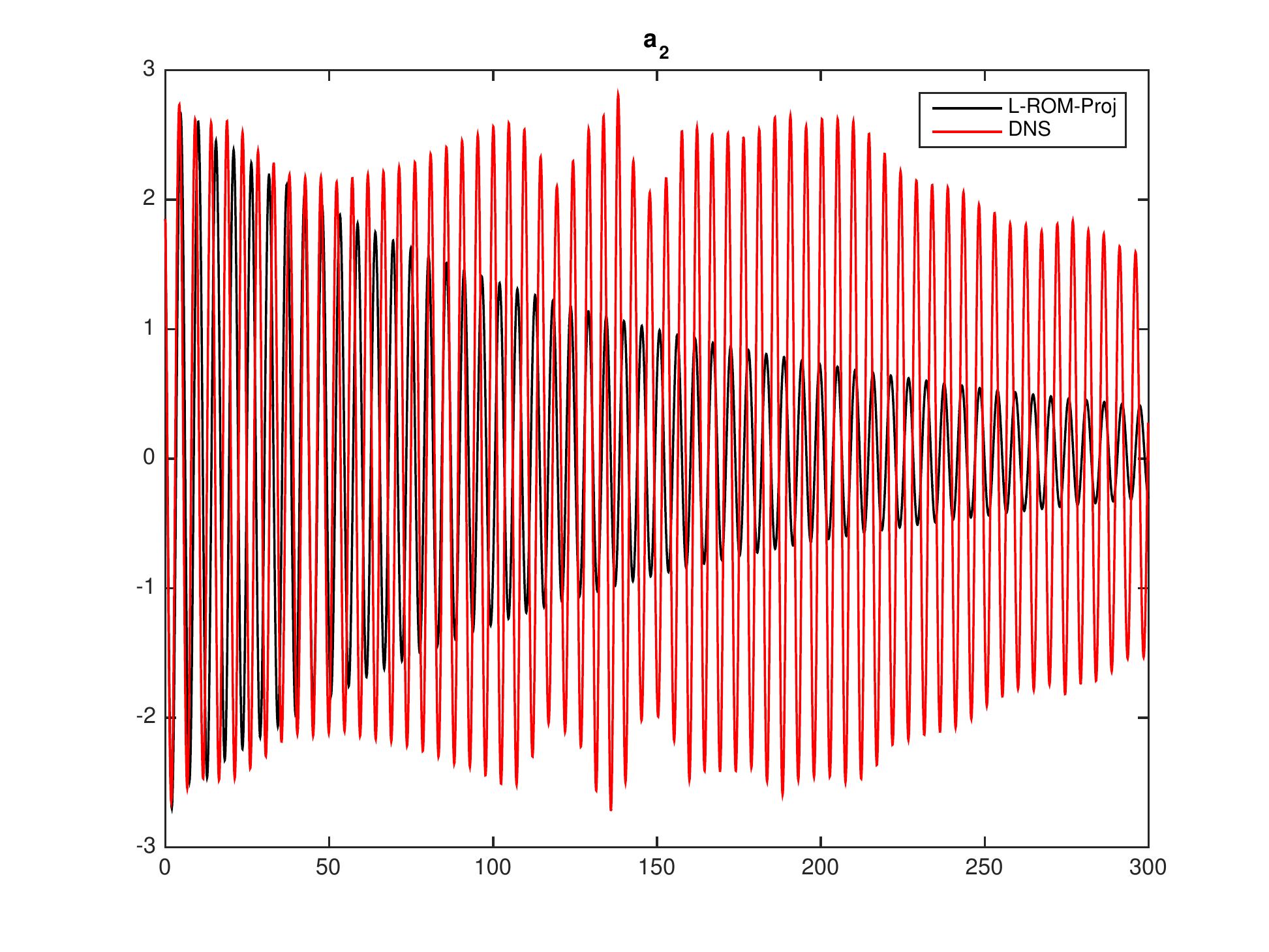}\end{minipage}
	\begin{minipage}[h]{0.4\linewidth} \includegraphics[width=0.9\textwidth]{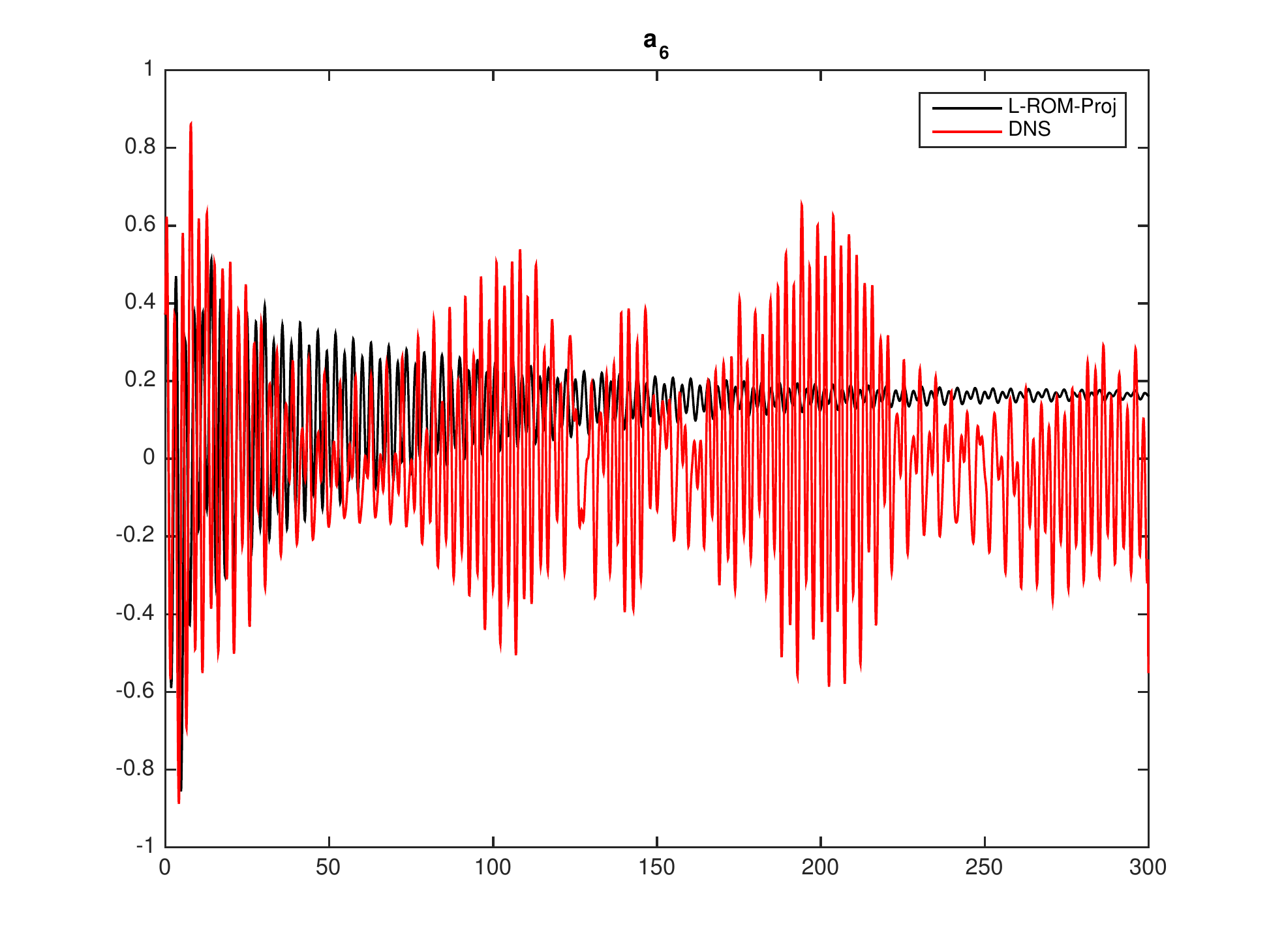}\end{minipage}\\
	\begin{minipage}[h]{0.4\linewidth} \includegraphics[width=0.9\textwidth]{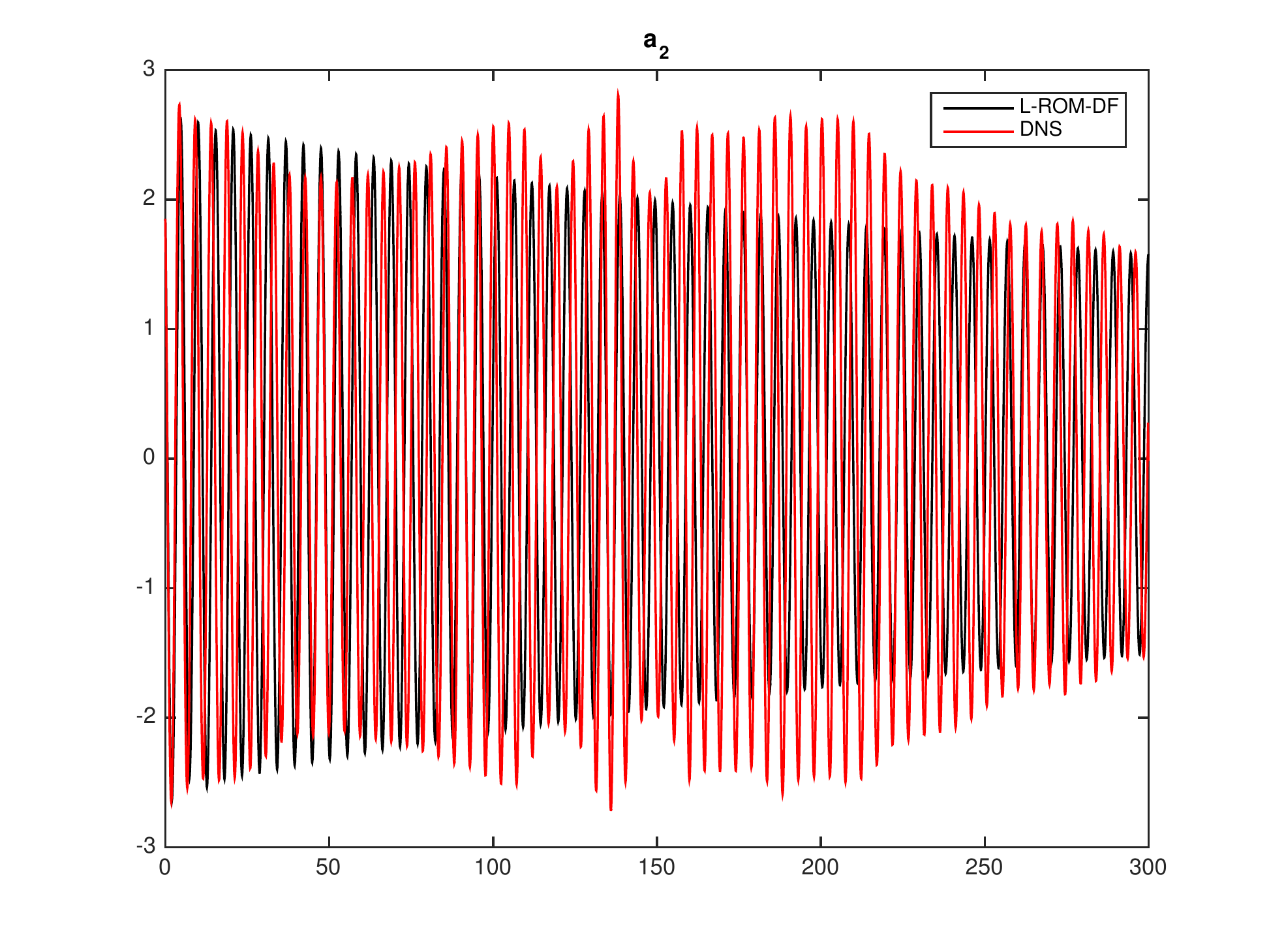}\end{minipage}
	\begin{minipage}[h]{0.4\linewidth} \includegraphics[width=0.9\textwidth]{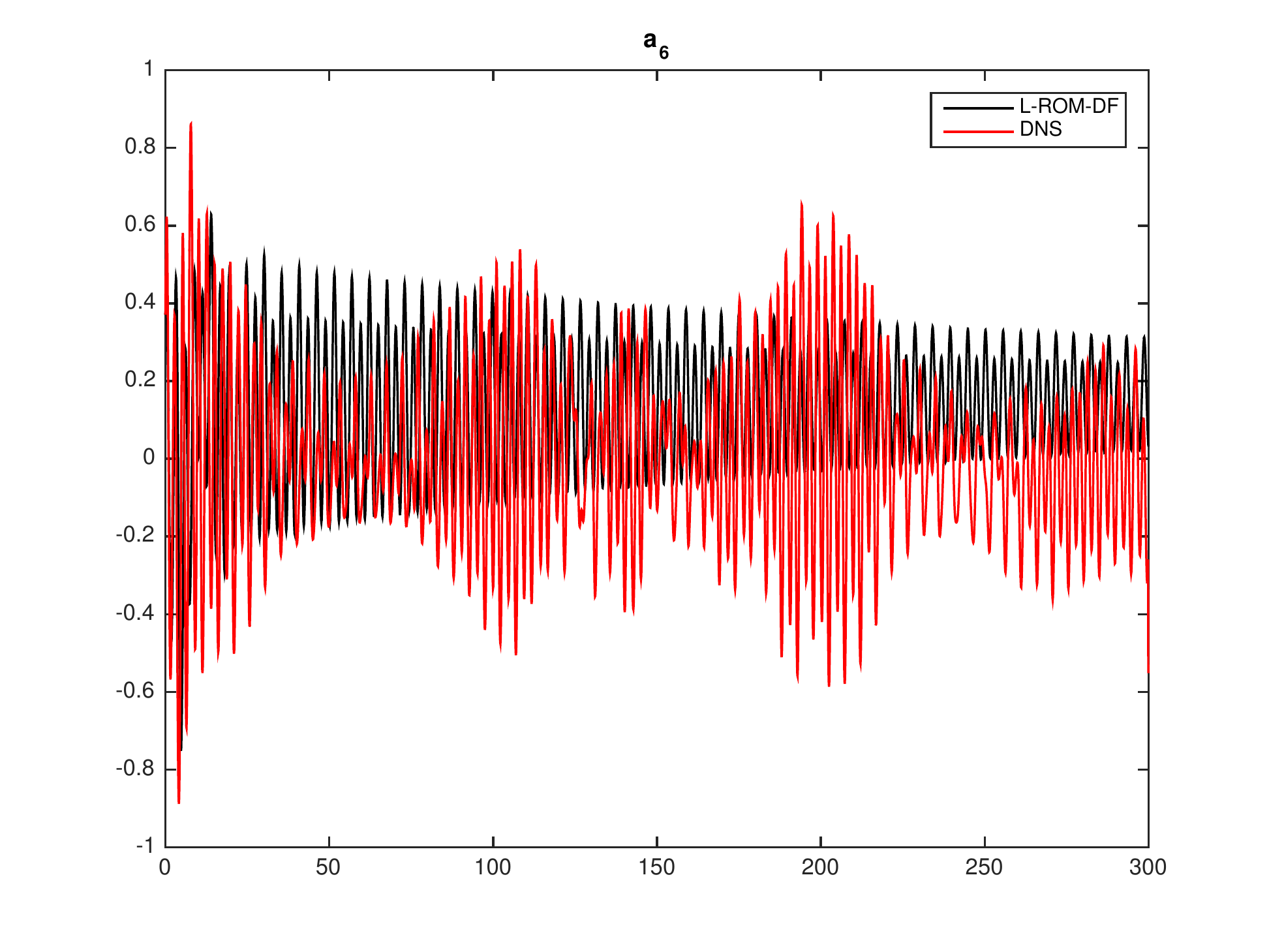}\end{minipage}\\
	\begin{minipage}[h]{0.4\linewidth} \includegraphics[width=0.9\textwidth]{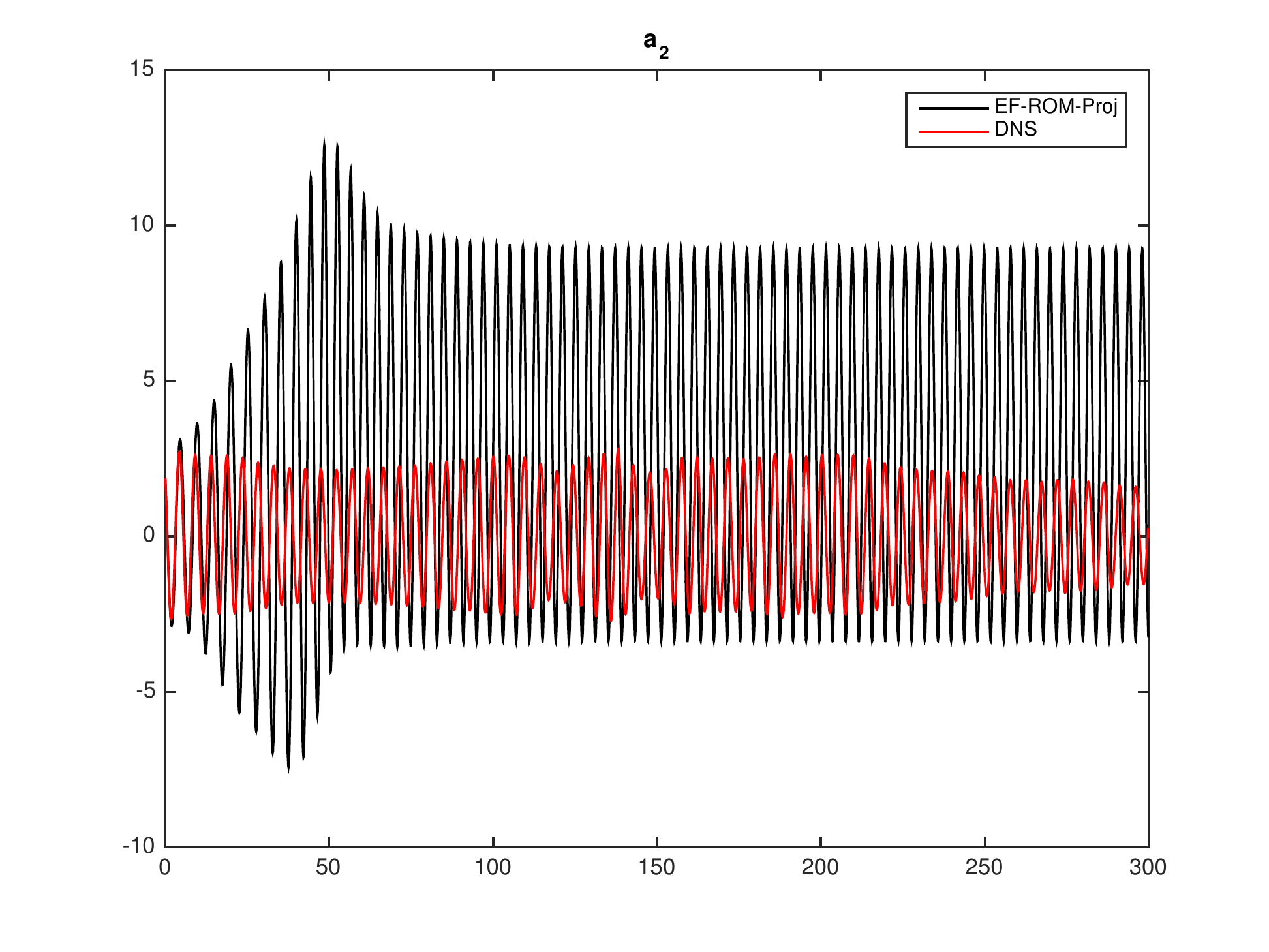}\end{minipage}
	\begin{minipage}[h]{0.4\linewidth} \includegraphics[width=0.9\textwidth]{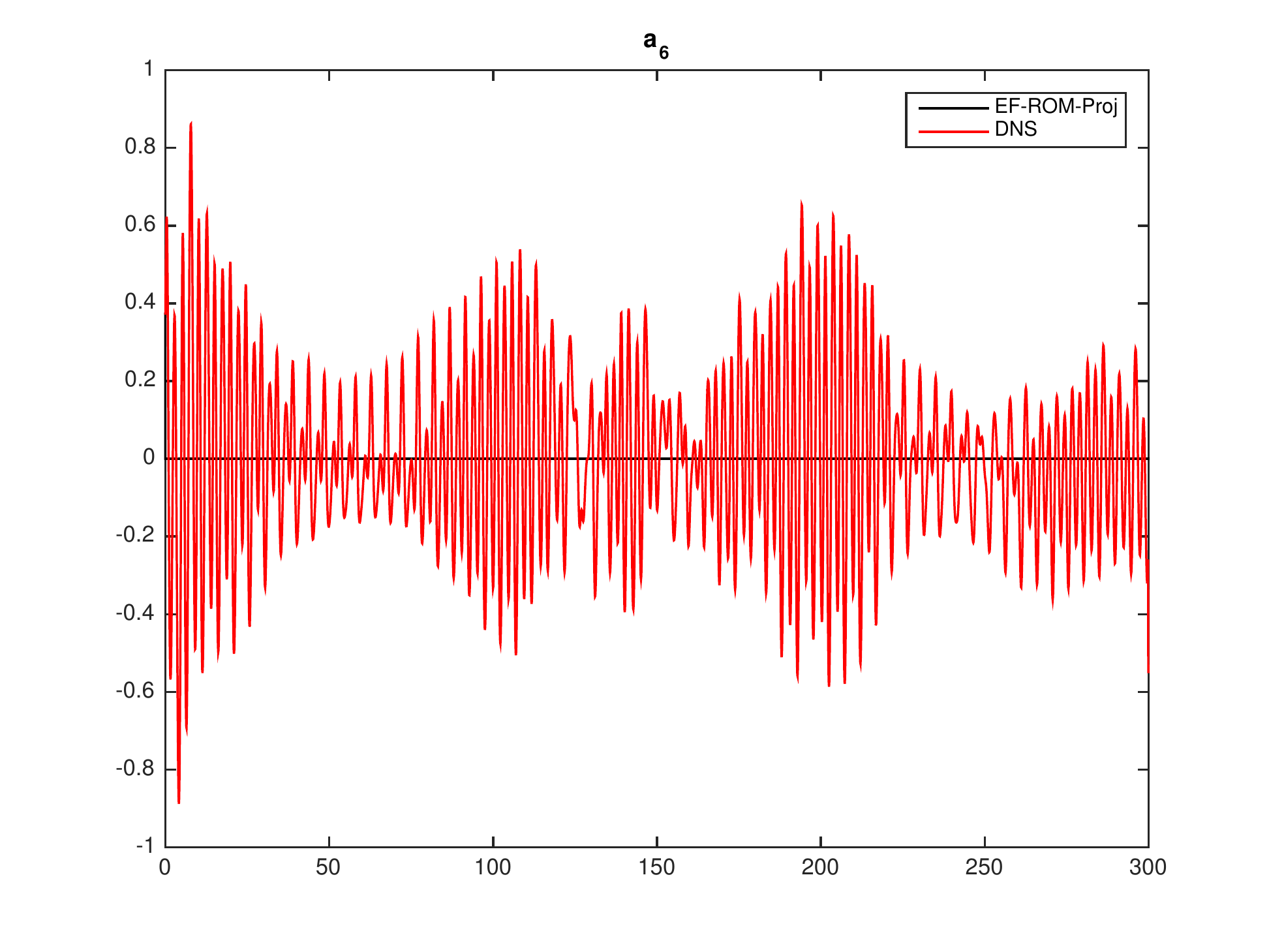}\end{minipage}\\
	\begin{minipage}[h]{0.4\linewidth} \includegraphics[width=0.9\textwidth]{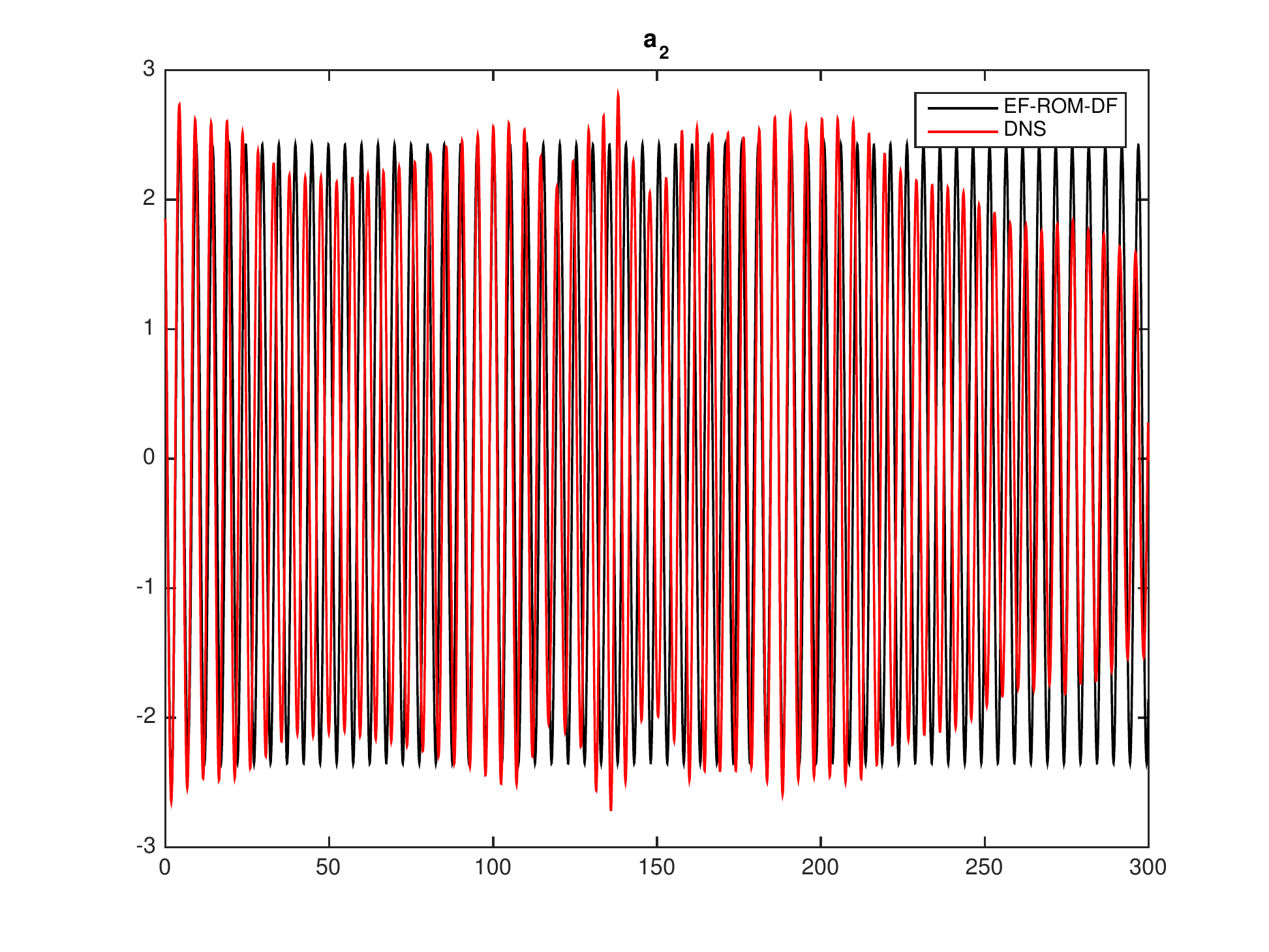}\end{minipage}
	\begin{minipage}[h]{0.4\linewidth} \includegraphics[width=0.9\textwidth]{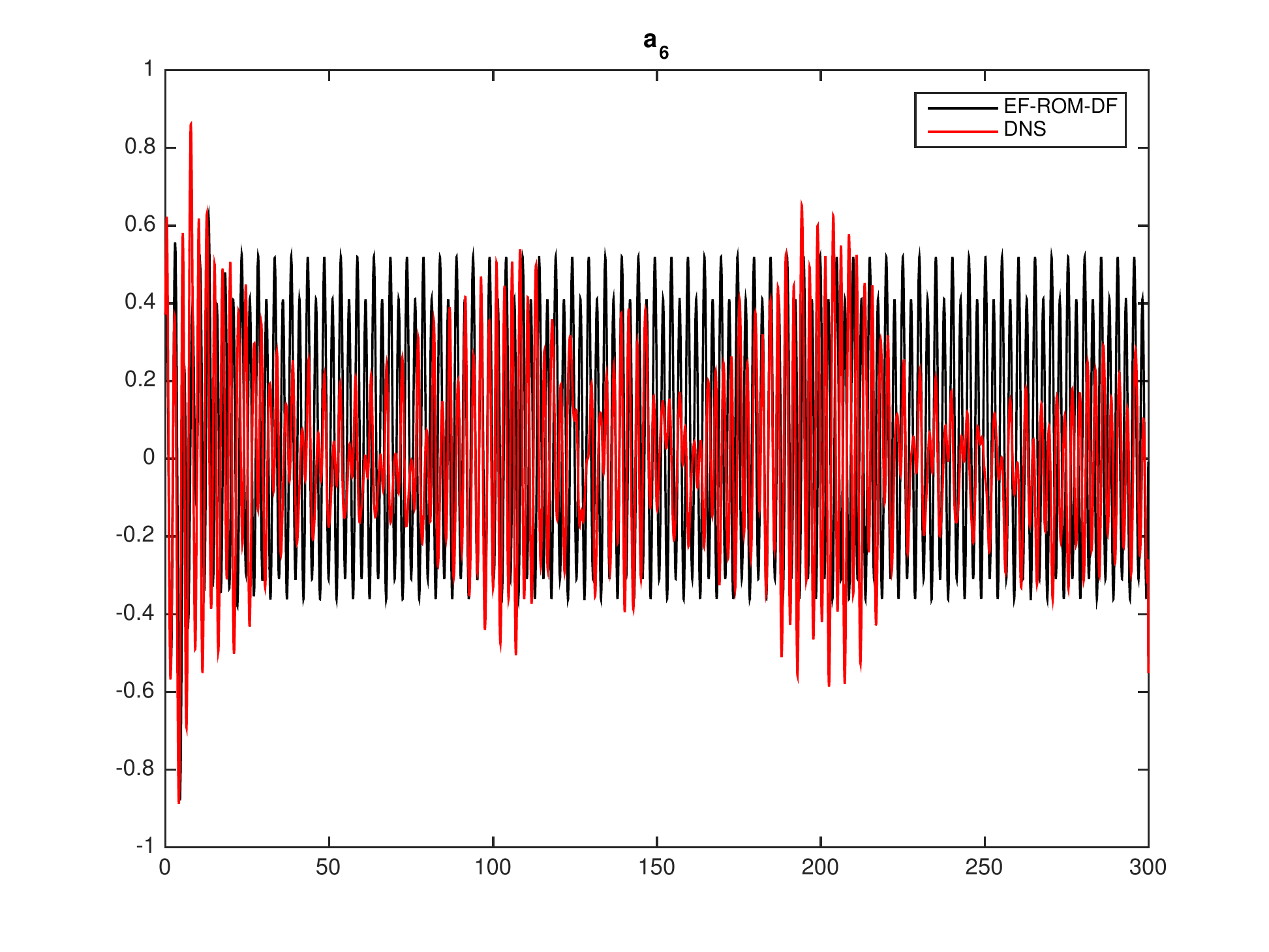}\end{minipage}\\
\end{figure}

Table~\ref{tab:summary} displays the ranking of the four Reg-ROM/filter combinations (i.e., L-ROM-DF, L-ROM-Proj, EF-ROM-DF and EF-ROM-Proj). 
The numbers in Table~\ref{tab:summary} represent the rank of the Reg-ROM/filter combinations (with $1$ the best and $4$ the worst). 
We emphasize that the results in Table~\ref{tab:summary} represent a general evaluation of the Reg-ROM/filter combinations for all criteria used.
The following overall rankings emerge: 
The EF-ROM-DF is the most accurate with respect to all the criteria  considered. 
Indeed, the EF-ROM-DF clearly yields:
(i) the best energy spectrum (see Fig.~\ref{fig:spectrum});
(ii) the best rms values, although L-ROM-DF was a close second (see Fig.~\ref{fig:rms}); and
(iii) clearly the best time evolutions of the POD coefficients $a_2$ and $a_6$ (see Fig.~\ref{fig:a2-a6}).
Furthermore, with respect to the other two criteria (the mean velocity components in Fig.~\ref{fig:mean} and the Reynolds stresses in Fig.~\ref{fig:stress}), the EF-ROM-DF performs at least as well as the other ROMs.
Thus, we conclude that the EF-ROM-DF yields the most accurate average and instantaneous numerical results.
The EF-ROM-Proj is consistently the least accurate. 
The L-ROM-DF and the L-ROM-Proj yield similar results, with an advantage for the former. 
Besides the rankings, Table~\ref{tab:summary} suggests that the spatial filter has a higher impact on the Reg-ROM than the regularization used.
Indeed, the DF generally yields better results than Proj for both the EF-ROM and L-ROM.
Finally, it should be emphasized that EF-ROM-DF, L-ROM-DF and L-ROM-proj performed significantly better than the standard G-ROM, whereas EF-ROM-Proj performed significantly worse than the G-ROM.

        \begin{table}[h]
            \centering
            \caption
              {
                Ranking of the L-ROM-DF, L-ROM-Proj, EF-ROM-DF and EF-ROM-Proj.
            }
            \label{tab:summary}
            \begin{tabular}{|c|c|c|c|}
                \hline
                & & & \\ [-0.2cm]
                L-ROM-DF & L-ROM-Proj & EF-ROM-DF & EF-ROM-Proj \\ [0.2cm]
                \hline
                & & & \\ [-0.2cm]
                2 & 3 & 1 & 4 \\ [0.2cm]
                \hline
            \end{tabular}
        \end{table}

\clearpage 

\section{Conclusions and Future Work}
    \label{sec:conclusions}
    
A new regularized ROM (Reg-ROM) was proposed: the evolve-then-filter ROM (EF-ROM).
The Leray ROM (L-ROM) was also investigated.
Both Reg-ROMs used explicit POD spatial filtering to regularize (smooth) some of the terms in the standard Galerkin ROM (G-ROM). 
Two explicit ROM spatial filters were investigated: a POD projection (Proj) on a subspace of the POD space and a POD differential filter (DF). 
To study the effect of the ROM spatial filtering on the Reg-ROMs, four Reg-ROM/filter combinations were considered: L-ROM-Proj, L-ROM-DF, EF-ROM-Proj and EF-ROM-DF. 
These four Reg-ROM/filter combinations were assessed in the numerical simulation of the 3D flow past a circular cylinder at $Re = 1000$. 
The Reg-ROM/filter combinations were tested with optimal values of $\delta$ (for the DF) and $r_1$ (for the Proj). 
These optimal values were determined by requiring that the DNS (benchmark) and Reg-ROM/filter combinations be as close as possible on the average. 
We emphasize, however, that the parameter values were optimized on a short time interval ($[0,75]$), whereas the ROMs were investigated on a significantly longer time interval ($[0,300]$).
The following criteria were used in the numerical assessment of the Reg-ROM/filter combinations: the kinetic energy spectrum, the mean velocity, the Reynolds stresses, the root mean square values of the velocity fluctuations, the time evolution of the POD coefficients, and the Strouhal number.
The numerical investigation of the four Reg-ROM/filter combinations yielded the following conclusions: 
(i) The EF-ROM-DF was clearly the most accurate, the EF-ROM-Proj was the least accurate, and the L-ROM-DF was more accurate than L-ROM-Proj.
We also note that the EF-ROM-DF, L-ROM-DF and L-ROM-Proj performed significantly better than the standard G-ROM. 
(ii) The explicit ROM spatial filter had a higher impact on the Reg-ROM than the regularization used.
Indeed, the DF generally yielded better results than Proj for both the EF-ROM and L-ROM.
(iii) The CPU times of all four Reg-ROM/filter combinations were orders of magnitude lower than the CPU time of the DNS.

These first steps in the investigation of the new EF-ROM yielded encouraging results. 
There are, however, several research directions that could be further pursued. 
For example, using an approach similar to that utilized in~\cite{ervin2012numerical} in a FE context could increase the EF-ROM accuracy by limiting its numerical dissipation.
Furthermore, although the same explicit Euler method was used in all the ROMs (which ensured a fair comparison), one could investigate more accurate numerical discretizations.
One could also test the new EF-ROM in the numerical simulation of more complex convection-dominated flows.
The preprocessing spatial filtering in~\cite{aradag2011filtered} should certainly be used to eliminate the noise in the snapshots (which, of course, is to be expected for the relatively coarse meshes utilized in complex applications). 
Although a necessary step in complex convection-dominated flows, this preprocessing spatial filtering will probably be not sufficient for ROMs that employ relatively few POD modes.
In this case, one could investigate whether the new regularized EF-ROM can further stabilize the numerical simulations and allow accurate approximations of complex convection-dominated flows.
Finally, one could perform a comparison of the new EF-ROM with other ROM stabilization strategies, such as those in~\cite{amsallem2012stabilization,balajewicz2013low,barone2009stable,bergmann2009enablers,carlberg2013gnat,kalashnikova2010stability,wang2012proper,xiao2014non}.
Of course, this would be a daunting task (which could possibly explain why it has not been performed so far); it would, however, provide invaluable insight to anyone who intends to use ROMs in {\it realistic} convection-dominated flows.

\clearpage

\bibliographystyle{plain}
\bibliography{traian}

\end{document}